# Directed Evolution of Microorganisms for Engineered Living Materials

Julie M. Laurent,[1] Ankit Jain,[2] Anton Kan,[1] * Mathias Steinacher,[1] Nadia Enrriquez Casimiro,[1] Stavros Stavrakis,[2] Andrew J. deMello,[2] André R. Studart [1] *

[1] Complex Materials, Department of Materials, ETH Zürich, 8093 Zürich, Switzerland

[2] Institute for Chemical and Bioengineering, Department of Chemistry and Applied Biosciences, ETH Zürich, 8093 Zürich, Switzerland

* corresponding authors

**Abstract**

Microorganisms can create engineered materials with exquisite structures and living functionalities. Although synthetic biology tools to genetically manipulate microorganisms continue to expand, the bottom-up rational design of engineered living materials still relies on prior knowledge of genotype-phenotype links for the function of interest. Here, we utilize a high-throughput directed evolution platform to enhance the fitness of whole microorganisms under selection pressure and identify novel genetic pathways to program the functionalities of engineered living materials. Using *Komagataeibacter sucrofermentans* as a model cellulose-producing microorganism, we show that our droplet-based microfluidic platform enables the directed evolution of these bacteria towards a small number of cellulose overproducers from an initial pool of 40'000 random mutants. Sequencing of the evolved strains reveals an unexpected link between the cellulose-forming ability of the bacteria and a gene encoding a protease complex responsible for protein turnover in the cell. The ability to enhance the fitness of microorganisms towards specific phenotypes and to discover new genotype-phenotype links makes this high-throughput directed evolution platform a promising tool for the development of the next generation of engineered living materials.

**Introduction**

Engineered living materials (ELMs) represent an enticing new class of materials whose structure and properties are governed by metabolically active microorganisms. [1-5] Such microorganisms can be temporarily used to generate intricate structures not accessible via conventional manufacturing methods and are often kept alive in the host structure to create a material with living, adaptive functionalities. The ability to grow, self-heal, remodel, and even make decisions are some of the unique adaptive functionalities displayed by engineered living materials. [6] Importantly, these functionalities are encoded in the DNA of wild-type microorganisms and may also be programmed using synthetic biology tools. [7-9] Besides their adaptive capabilities, engineered living materials are usually grown in aqueous media under ambient temperatures and pressures, thus offering a more energetically efficient alternative to their synthetic counterparts. Moreover, microorganisms rely on abundant and environmentally friendly chemistries to build materials, which makes ELMs more sustainable and resource-efficient compared to artificial materials. In addition, the rich biochemical diversity in microorganisms provides an immense design space for the engineering of adaptable living materials.

To achieve functionalities of relevance in engineering applications, living microorganisms must have evolved traits to fulfill a biological function that matches an engineering need. Living materials containing self-regenerating mycelia [6] and bio-cementing bacteria [10-12] are illustrative examples in which wild-type species can be selected for an engineering task based on their biological function in the natural environment. Another approach is to re-program the genome of microorganisms with synthetic biology tools to achieve functionalities desired in an engineering context. This compelling strategy has been employed to create genetically engineered living materials for catalysis, energy conversion, sensing, electronic, and biomedical applications. [2,8,9,13-21] Despite the demonstrated success of these strategies, a mismatch often exists between the biological function of wild-type species and the desired engineering functionalities. When trying to engineer microorganisms, the concerted action of many genes involved in the envisioned functionalities may complicate the process, and trade-offs might need to be considered. In other instances, the genes encoding for a specific function are also unknown, thus preventing a clear connection between genotype and phenotype. Even if a function-encoding gene is known, the fitness of the whole microorganism to an engineering setting will likely involve more than one specific trait. These challenges call for the development of other strategies for the selection of genetically programmable microorganisms for engineered living materials. The directed evolution of whole microorganisms is a promising route to fill this gap.

Directed evolution has been widely used to improve the selectivity and activity of enzymes by exploring the vast design space available in the genome of microorganisms. [22-24] Following the principles of natural selection, this approach accelerates the iterative process of genetic diversification and selection for a specific desired phenotype. Gene diversification typically occurs by inducing targeted or random mutations in the genome of the microorganism that produces the enzyme of interest. [25-27] This results in a library of distinct mutants that are afterward screened based on the performance of the enzyme. Since the probability of finding a mutant with improved performance is very low, directed evolution of enzymes is often performed with an initial library containing $10^4$ to $10^7$ mutants. [24,28-30] Rare high-performance mutants can be selected from this vast pool using microfluidic approaches, which can operate at extremely high throughput if individual assays are performed inside droplets. [24] While a phenotypic selection approach has previously been applied to screen magnetotactic bacteria mutants [30] high-throughput droplet-based directed evolution tools have not yet been used to select microorganisms and discover apparently unconnected genotype-phenotype links useful for engineered living materials.

Here, we utilize a droplet-based microfluidic platform for the directed evolution of cellulose-producing bacteria that can be integrated into complex engineered living materials. Given the broad interest in cellulose fibers as a sustainable natural resource, the bacterium *Komagataeibacter sucrofermentans*, known as a high-yield cellulose producer, was chosen to illustrate the platform and to shed light on genotype-phenotype links in this microbial species. Bacteria were evolved towards an overproducer phenotype by encapsulating, incubating, and sorting single bacteria mutants within microfluidic droplets. Engineered living materials with complex shapes were manufactured by 3D printing gels containing the evolved bacteria. Finally, the sorted overproducers were compared with the wild-type strain in terms of cellulose bulk production and genetic sequence to elucidate possible mechanisms controlling cellulose bio-synthesis in bacteria.

**Results and Discussion**

*K. sucrofermentans* is a gram-negative aerobic bacterium that produces long cellulose nanofibers found in food products, wound dressing materials, and high-end acoustic membranes. [31] This microorganism metabolizes sugars to UDP-glucose, which is then catalyzed by the transmembrane cellulose synthase complex into β(1-4) glucan chains that self-assemble into cellulose fibers while being exported through the cell wall (Figure 1a). The core cellulose synthase machinery in *K. sucrofermentans* is encoded by multiple copies of genes *bcsA*, *bcsB*, *bcsC*, and *bcsD*, alongside several accessory genes, which are present in various operons throughout the genome. [32,33] Since the function of some of those genes is not yet fully understood and that other genes are expected to be involved in the regulation of the cellulose formation process, deliberate genome editing in this microorganism is not straightforward. To circumvent this issue, we opted to use whole-genome mutagenesis in our directed evolution approach to generate variants with desirable cellulose synthesis phenotypes without prior assumptions on genotype-phenotype links.

For many established and prospective applications, the slow growth of bacteria-produced cellulose pellicles represents a challenge that prevents the broader utilization of this sustainable biopolymer. To address this issue, we established the amount of cellulose produced by the bacteria in 24 hours as the target phenotype in our directed evolution process. Our hypothesis is that bacteria subjected to this selection pressure will evolve strategies to increase the throughput of their biological machinery (Figure 1b). If successful, such an evolutionary process will lead to strains with genetic mutations that favor faster cellulose production. These evolved microorganisms can then be exploited to create cellulose-based macroscopic objects using state-of-the-art manufacturing technologies, such as casting, molding, and 3D printing. By combining the bottom-up cellulose-producing capability of the evolved bacteria with the top-down shaping freedom of conventional manufacturing, we expect to generate engineered living materials with unique multiscale architectures (Figure 1c).

The directed evolution process of our cellulose-producing microorganism starts by creating a library of mutants (Figure 1d). In contrast to the directed evolution of enzymes, our goal is to enhance bacterial cellulose production by exploring the entire genome of the microorganism rather than introducing mutations only in the genes encoding a specific target protein. To this end, we selected a random mutagenesis approach using UV-C light to generate the mutant library. This approach allows us to tap into the broad diversity of possible genetic mutants without prior assumptions about genes that might control cellulose biosynthesis. From this initial library, we selected mutants with the desired phenotype through a series of directed evolution cycles using a droplet microfluidic platform (Figure 1e-g). The encapsulation of a single bacterium in individual droplets is a crucial feature of this approach, since it establishes a direct link between genotype and phenotype.

The microfluidic platform consists of three parts: a droplet generator for cell encapsulation, an incubation container for cellulose production and cell growth in droplets, and a high-speed fluorescence-activated droplet sorter (FADS). [34] To enable the directed evolution of individual cells with unique genomes, we encapsulate one microorganism per droplet by adjusting the initial cell concentration in the culture medium containing a cellulose-binding fluorescent dye. After encapsulation, cell-laden stable droplets are incubated in glass containers for up to 4 days under optimum growth conditions (Figure S1, Supporting Information). Droplets are then reinjected in the FADS and those containing mutants that overproduce cellulose are dielectrophoretically separated from the rest of the population by applying an electrical pulse when the measured fluorescence exceeds a user-defined threshold. This approach relies on the fact that the amount of cellulose produced by the encapsulated bacterium is directly correlated to the measured fluorescence. Microorganisms from sorted droplets are finally compared to the native strain to identify true overproducers in larger cultures.

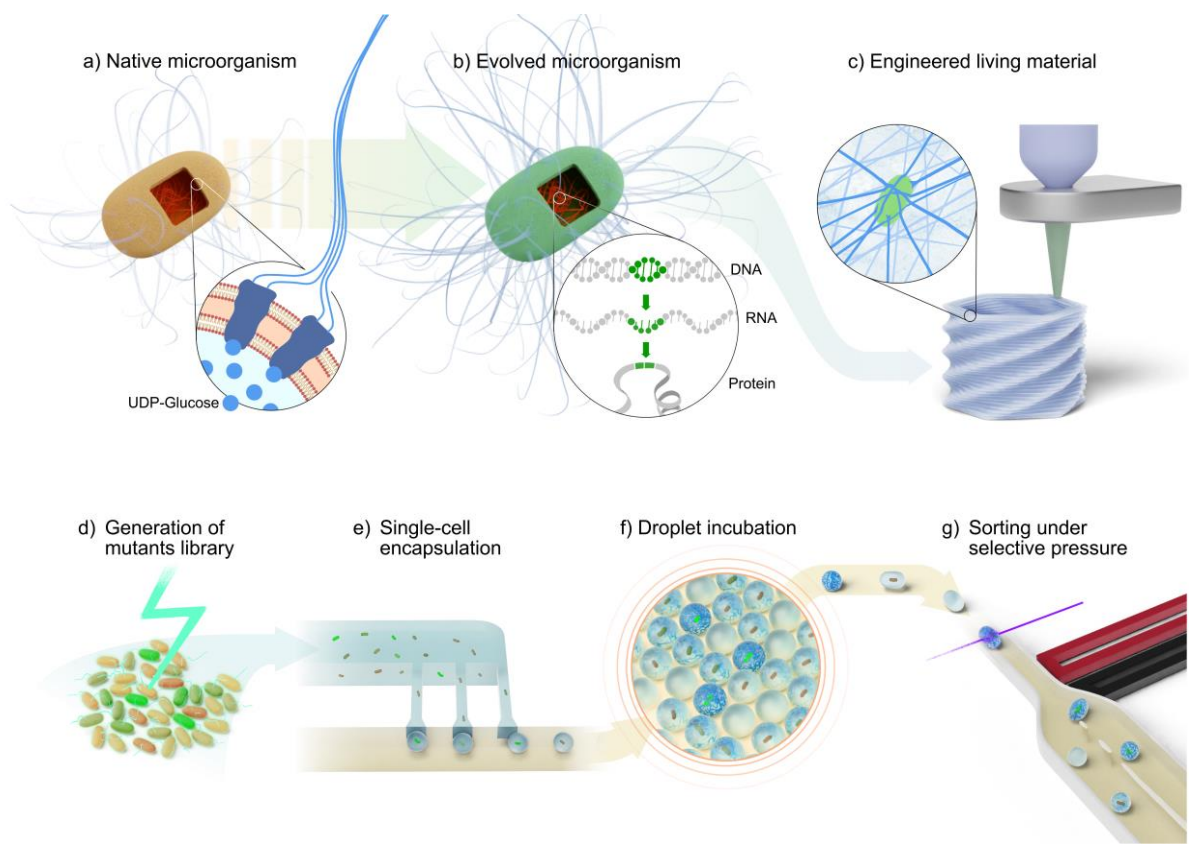

**Figure 1. Direct evolution of cellulose-producing microorganisms for engineered living materials. a.** Native *K. sucrofermentans* metabolizes sugars to UDP-glucose, which is polymerized into β(1-4) glucan chains catalyzed by the cellulose synthase transmembrane complex. The glucan chains are exported through the cell wall and self-assembled into bundles to form nanofibers of cellulose. **b.** Microorganisms evolved under selective pressure produce more cellulose, which can be due to mutations in their genome. Genetic mutations on a coding sequence translate into a change in the corresponding proteins, which could ultimately enhance cellulose production. **c.** Evolved cellulose overproducers can be shaped into 3D living objects by combining the bottom-up self-assembly of cellulose fibers with the top-down shaping freedom of state-of-the-art manufacturing. **d.** The directed evolution process of a natural microorganism starts with the generation of the library of mutants using whole-genome mutagenesis. **e.** Single cells are encapsulated in monodisperse droplets in the presence of a cellulose-binding fluorescent dye using a high-throughput microfluidic device. **f.** Off-chip incubation of cell-laden droplets allows for cellulose production and cell growth. **g.** Cellulose overproducers are selected using electrically controlled dielectrophoretic forces via fluorescence-activated droplet sorting (FADS).

The generation of bacterial mutants relies on the damage of DNA upon exposure of cells to UV-C irradiation. However, if DNA damage is too severe, the viability of cells and thus the directed evolution process are compromised. To create a large library of bacteria with a fraction of viable mutants, we investigated the effect of UV-C irradiation dose on the survival rate of cells (Figure 2a-c). Here, *K. sucrofermentans* suspended in a salt solution were irradiated with UV-C light at doses between 0.5 and 100 mJ/cm$^2$. After exposure, samples were kept in the dark for 1 hour to inhibit the onset of DNA repair mechanisms that prevent mutagenesis. [35] This was followed by the recovery of the cells in a nutrient-rich media for an hour (Figure 2a). To quantify the cell survival rate, bacterial cultures exposed to different light doses were directly frozen for subsequent colony-forming unit (CFU) counting analysis.

The CFU counting analysis revealed that the survival rate of the irradiated bacteria decreased from 98% to 17% upon an increase in the UV-C dose from 0.5 to 10 mJ/cm$^2$ (Figure 2b). The survival rate was calculated relative to a control sample, which underwent the same processing but was not irradiated. Exposure to the highest dose of 100 mJ/cm$^2$ led to severe DNA damage and complete cell death. Based on these results, a dose of 10 mJ/cm$^2$ was found to be most appropriate to ensure a high fraction of mutants in the population whilst keeping bacteria viable. Cells exposed to 10 mJ/cm$^2$ were further recovered for 50 hours in rich culture media as a primary selection step to enrich the suspension with fast-growing bacteria. During this period, the concentration of mutant bacteria was found to increase by 8-fold from 4.7 to 38.7 million CFU/ml (Figure 2c). This relatively concentrated cell suspension was frozen for later use as feedstock for the microfluidic encapsulation process.

Mutant and native bacteria were encapsulated in 49-μm droplets using a microfluidic device via the parallelized step emulsification approach (Figure 2d-e). Operation at a throughput of thousands of droplets per second and a droplet polydispersity index (PDI) of only 10$^{-3}$ allows the generation of large libraries of mutants with tightly controlled growth conditions that are ideal for directed evolution experiments. In the emulsification process, droplets of the diluted bacteria suspension are emulsified in a fluorocarbon oil and stabilized by a biocompatible surfactant. The number of cells encapsulated in a single microfluidic droplet is known to follow Poisson statistics and depends on the concentration of bacteria in the feedstock suspension. [36] We thus tuned the bacteria suspension concentration ($\lambda$ ~0.1 CFU/droplet) to generate a droplet population with ~90.5% of them being empty, ~9.0% containing a single cell, and ~0.5% containing more than 1 cell (Figure S2, Supporting Information). [24] In this way, cell-laden droplets should almost exclusively contain a single genome, coupling the genotype and phenotype.

A critical aspect of droplet-based directed evolution processes is to design an assay for fast on-chip quantification of the performance of encapsulated mutants. Leveraging the ease and availability of fluorescent-based detection tools, we used the Fluorescent Brightener 28 to quantify the production of cellulose by encapsulated bacteria. This commercially available dye is known to selectively bind to the (1-4)β bonds present in cellulose. To detect cellulose formation inside the droplets during incubation, we introduced the dye directly into the bacteria suspension used in the encapsulation process. The cellulose-forming capabilities of the encapsulated bacteria were measured using confocal microscopy after incubating the monodisperse droplets at 28°C for periods up to 4 days (Figure 2f and Figure S2, Supporting Information).

Confocal microscopy images acquired during the incubation period show that 6 hours were already sufficient for bacteria to start producing cellulose inside droplets. This indicates that the confinement in droplets does not stop the microorganism from synthesizing cellulose. At this early stage, the cellulose predominantly formed around the encapsulated bacteria. After extending the incubation period to 24 hours, the cellulose produced was no longer confined to the vicinity of the bacteria but occupied most of the droplet volume. Fluorescence distribution histograms obtained by image analysis clearly show the emergence of a population of cellulose-containing droplets after 6 and 24 hours of incubation (Figure 2g-i). These results validate the effectiveness of the fluorescence-based approach to quantify the cellulose-producing capabilities of the bacteria encapsulated in droplets.

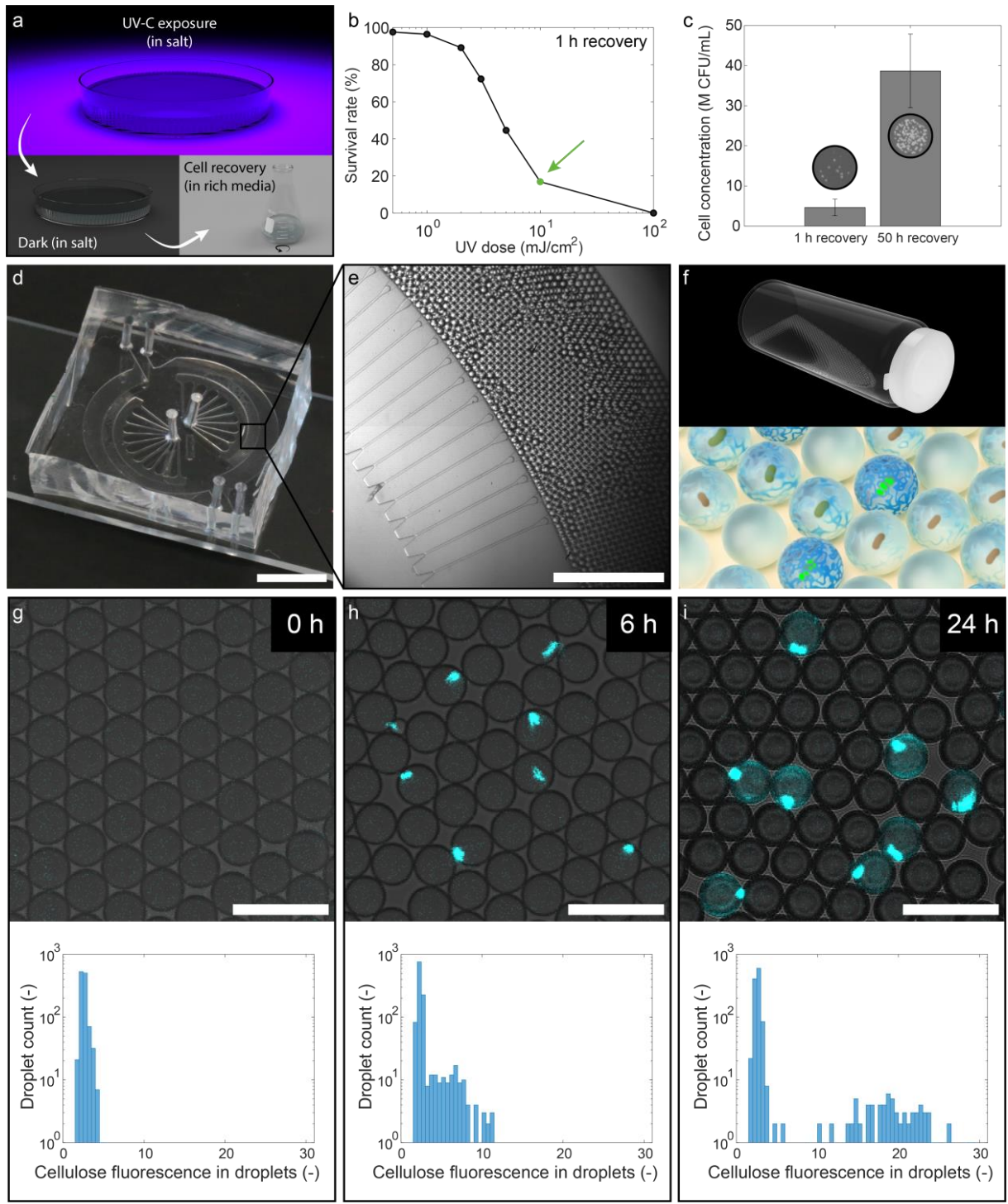

**Figure 2. Mutagenesis, single-cell encapsulation, and bacterial cellulose quantification in droplets. a.** Mutagenesis steps used to create a mutant library. After UV-C exposure, the cell suspension in salt is left in the dark for 1 h. This is followed by a recovery step in rich media for 1 h or 50 h, and finally freezing. **b.** Survival rates of cell suspensions exposed to different UV-C doses compared to a non-exposed sample. The culture exposed to 10 mJ/cm$^2$ has a survival rate of 17% and was selected for the directed evolution process (green arrow). **c.** Impact of cell recovery step after mutagenesis to enrich the suspension with fast-growing bacteria. Error bars represent the standard deviation. Insets show the colony forming units (CFUs) from 5 µL cell suspensions at a 10$^{-3}$ dilution after 1 and 50 h of recovery. **d.** Step emulsification microfluidic device used to generate populations of monodisperse droplets. Scale bar: 1 cm. **e.** Parallelized channels producing monodisperse droplets at a rate of 2000 droplets per second. Scale bar: 1 mm. **f.** Horizontal droplet incubation vial to ensure a droplet monolayer and therefore equal oxygen availability to encapsulated bacteria (top). Over one day

of incubation, cells grow and cellulose is produced in the droplets (bottom). **g-i.** Single-cell laden droplets over different incubation times (λ ~0.1 CFU/droplet). Native bacteria were used in these experiments to assess cellulose formation in droplets. The top row displays representative confocal microscopy images, with droplets shown in grey and fluorescently labeled cellulose in cyan. Scale bar: 100 µm. The bottom row presents corresponding histograms of cellulose fluorescence per droplet, quantified by image analysis (*n* ~1200 droplets). First, cellulose forms predominantly around the encapsulated bacteria (**h**), and then occupies most of the droplet volume (**i**).

Due to the high emulsification rate of 2000 droplets per second, the microfluidic platform enabled the generation of a library of 1.2 million droplets in only 10 minutes. This corresponds to approximately 100'000 single-cell encapsulated mutants. A fraction of this large pool of mutants (430'000 droplets, ~40'000 mutants) was screened in 9 minutes in the microfluidic droplet sorter to select for potential cellulose overproducers after 24 hours of droplet incubation off-chip (800 Hz). In such a FADS device, the flowing droplets are excited using a laser, and the emitted fluorescence is detected by a photomultiplier tube (PMT; Figure 3a). [24] The PMT voltage signal is used as a proxy for the fluorescence quantification in droplets. In addition to the cellulose-binding fluorescent dye, fluorescein was added in the droplets to serve as a baseline detection for all droplets.

Droplets with fluorescence above a user-defined threshold corresponding to cellulose overproducers are dielectrophoretically pulled into a separate outlet on the chip, while the vast majority is discarded in a waste outlet (Figure 3a,b). Preliminary experiments showed the importance of using a recovery time of 50 hours after UV-C exposure to obtain a strong cellulose fluorescence signal after 1 day of bacteria incubation in droplets (Figure S3, Supporting Information). For droplets loaded with bacteria obtained after the 50-hour recovery process, the PMT voltage threshold was set to 2.75 V to separate the top 1.25% most fluorescent mutants from the pool (Figure 3c).

Sorted droplets contauining evolved bacteria were collected and spread on solid media to enable the growth of individual mutants into colonies (Figure S4, Supporting Information). Five colonies of evolved bacteria (Ev1 to Ev5) were arbitrarily chosen for further analysis. The evolved bacteria were compared to the native and control (0 mJ/cm$^2$) strains by measuring their cellulose-forming capabilities inside droplets or in bulk pellicles. For analysis in droplets, bacteria were re-encapsulated (λ ~0.1 CFU/droplet), incubated for 24 hours, and screened using the same microfluidic platform used for the initial evolution experiment. This time, no sorting was performed. Instead, we focused on the analysis of the fluorescence intensity distributions obtained for the native, control, and evolved bacteria.

Evolved mutants clearly outperformed the native and control bacteria in terms of cellulose production inside droplets. Fluorescence intensity histograms obtained for these three variants show a Gaussian-like distribution that is strongly skewed towards high fluorescence values (Figure 3d). To quantify the performance of the distinct bacteria, we fitted the distributions to a Gaussian mixture function, taking into account the baseline corresponding to cell-empty droplets. The mean fluorescence intensity of the peak corresponding to droplets containing cellulose could thus be compared between strains. The control and evolved bacteria displayed a mean cellulose fluorescence 36% and 52%, respectively, higher than that measured for the native strain. This enhanced cellulose-producing capability is also evidenced by higher maximum fluorescence signals from droplets loaded with evolved bacteria. Indeed, the percentage of variants with a fluorescence higher than 3 V was 0.07%, 0.84%, and 2.46% for the native, control, and evolved samples, respectively.

Importantly, the over-production of cellulose by the evolved bacteria is not limited to droplets but can also be translated into the bulk manufacturing of thicker cellulose pellicles. To illustrate this, we cultivated the evolved, native, and control microorganisms from single colonies in liquid culture that allows for the growth of a bacterial cellulose (BC) pellicle at the air-water interface in static conditions (Figure 3e). A long incubation time of 12 days was chosen to ensure that cells reached the stationary growth phase and to maximize cellulose production. This also minimized the possible effect of slight differences in the initial inoculation amounts (Figure S5, Supporting Information). Pellicles were then washed and dried to quantify the amount of cellulose produced by the different variants by visual inspection, weight measurements, and thermogravimetric analysis (TGA).

Visual inspection of freeze-dried pellicles clearly shows the production of thicker pellicles by the evolved bacteria compared to the native species (Figure 3f). Both scale measurements and thermogravimetric analysis of air-dried pellicles heated up to 650°C (Figure S6, Supporting Information) revealed that 4 of the 5 evolved strains chosen for analysis produced 54-70% more cellulose compared to the native bacteria (Figure 3g). The long fibers obtained from these strains show the typical interwoven microstructure of bacterial cellulose (Figure S7, Supporting Information). Interestingly, the non-sorted control strain also showed a 19% increase in cellulose production relative to the native species, suggesting that the stresses imposed on the bacteria during the treatments for mutagenesis (exposure to salt, darkness, and the recovery process) can promote cellulose formation. These results indicate that mutants selected for fast cellulose production in droplets also led to bulk pellicles with the highest total cellulose mass after 12 days of incubation.

To evaluate the stability of the enhanced phenotype, we performed passage experiments in which bacterial cellulose pellicles were grown from previous pellicles over multiple generations. The results showed that the higher production of cellulose by the selected evolved strain (Ev5) was stable even after 5 passages (Figure S8, Supporting Information). Besides this stability experiment, we also performed an additional directed evolution cycle of the Ev5 strain to explore the possibility of further phenotype improvement. To this end, the evolved strain was mutated, encapsulated, incubated, and sorted with two different thresholds: the same as in the first round (2.75 V) and a more stringent one (3.60 V). The selected strains of this second round of directed evolution did not show a further increase in cellulose production but maintained a similar production as in the first evolution round (Figure S9, Supporting Information). Additional mutations leading to further cellulose production increase might be possible, but it would require more directed evolution cycles.

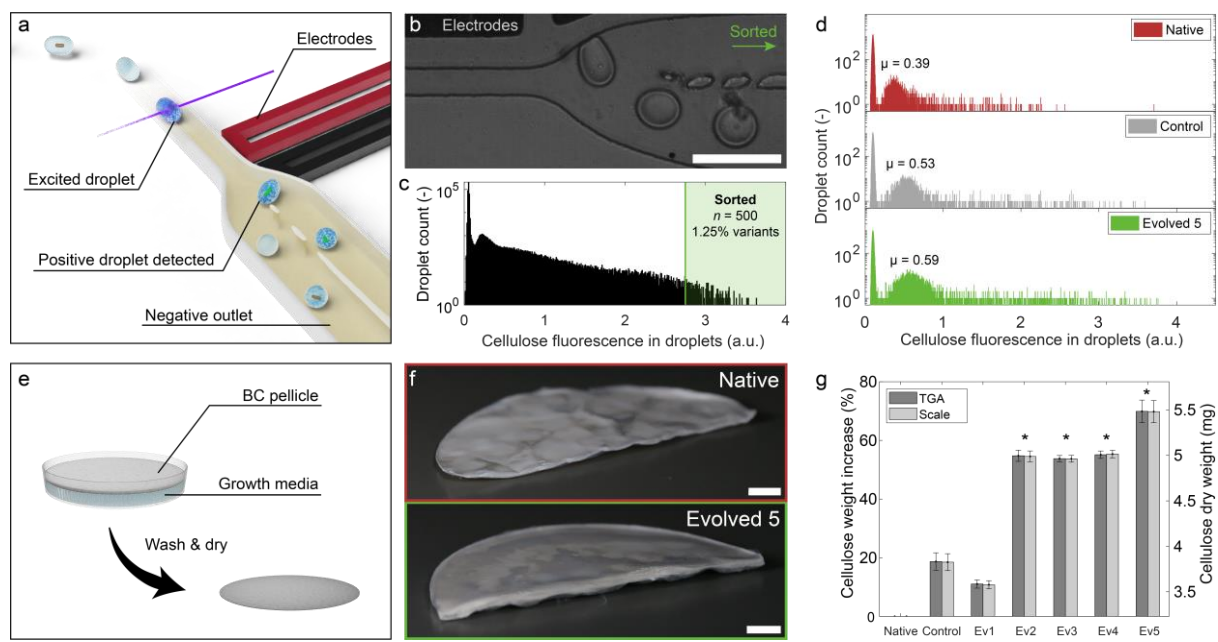

**Figure 3. Fluorescence-activated droplet sorting and bacterial cellulose production by the evolved strain. a.** Schematics depicting the droplet sorting process. After 24 h of incubation, cell-laden (and empty) droplets pass through a laser, which excites the cellulose-bound fluorescent dye. If the PMT detector records a voltage higher than a user-defined threshold, the corresponding droplet is sorted dielectrophoretically. **b.** Image displaying the microfluidic sorter in action. Scale bar: 100 μm. **c.** Histogram of the cellulose fluorescence signals in droplets containing mutated cells (10 mJ/cm$^2$, 50 h recovery) after incubation for 24 h (λ ~0.1 CFU/droplet, 430'000 droplet events, ~40'000 mutants screened). Droplets with a cellulose fluorescence signal corresponding to a PMT voltage above 2.75 (green box) were sorted. The sorted droplets correspond to 0.12% of all droplets and ~1.25% of all mutant-laden droplets. **d.** Histograms of cellulose fluorescence signal in droplets containing native, control, or evolved cells (λ ~0.1 CFU/droplet, 15'000 droplet events). The fluorescence peaks were fitted with a Gaussian distribution, leading to mean cellulose-fluorescence values ($\mu$) of 0.388, 0.527, and 0.590 for the native, control, and evolved bacteria, respectively. **e.** For the formation of bulk materials, bacterial cellulose pellicles are grown from single colonies in media for 12 days, washed, and dried. **f.** Pictures of washed and freeze-dried bacterial cellulose pellicles (Ø = 9.65 cm, 100 mL 2X growth media). Scale bar: 1 cm. **g.** Weight measurements of washed and air-dried bacterial cellulose pellicles grown for 12 days from single colonies (Ø = 3 cm, 5 mL growth media). Weights were measured with both a laboratory scale and via TGA. Error bars correspond to the propagated error for the percentage data. Evolved strains (Ev2-5) showed significantly increased cellulose production (54-70%) compared to the native strain (*$p$ < 0.05, $n$ = 3).

In addition to forming thick bulk pellicles, the ability of the evolved bacteria to overproduce cellulose also opens new opportunities for the manufacturing of complex-shaped engineered living materials. To demonstrate this, we prepared a 3D printable gel that can be loaded with the evolved cellulose-producing bacteria (Figure 4a-c). Networking-forming silica particles, hyaluronic acid, and κ-carrageenan were used to tune the rheological properties of the gel. [3] Oscillatory rheology experiments confirmed that both cell-laden and cell-free gels display an elastic response at low shear stresses and become a viscous fluid above a well-defined yield stress of 200 Pa (Figure 4c and Figure S10, Supporting Information). These properties fulfill the rheological requirements for 3D printing via the extrusion-based direct ink writing (DIW) technique (Figure 4a). [37,38]

Cell-laden gels were successfully printed into a three-dimensional object with complex geometry on the centimeter scale (Figure 4b). In addition to rheology modifiers, the ink also contained nutrients required for the proliferation and growth of the embedded bacteria. This allowed for the in-situ production of cellulose fibers within the printed gel during an incubation period of 1 day. To gain insights into the effect of the bacterial strain on cellulose formation within the gel, we printed 12-mm discs containing the cellulose-binding dye, loaded them with either the evolved or native strains, and imaged them using confocal microscopy (Figure 4d-h). Fluorescence emitted by the cellulose-binding dye was taken as a proxy for the local concentration of cellulose produced by the bacteria. To restrict oxygen supply to the edges of the sample, the printed discs remained enclosed between a Petri dish and a cover slip during the incubation and imaging processes (Figure 4d).

Confocal images of the printed disks showed cellulose production predominantly along the edges of both types of samples next to the air-water interface. Notably, the evolved bacteria produced more cellulose than the native bacteria, leading to the formation of a cellulose-rich narrower ring at the edge of the disk (Figure 4f and Figure S11, Supporting Information). This contrasts with the broader and less intense fluorescent cellulose ring in the samples containing the native microorganisms (Figure 4e and Figure S11, Supporting Information). Radially integrated fluorescence measurements of the samples provide clear evidence of the distinct cellulose patterns created by the two bacterial strains (Figure 4g,h). This example illustrates how the architecture of engineered living materials can be finely tuned at various length scales by combining the top-down manufacturing capabilities of 3D printing with the bottom-up self-assembly processes controlled by different strains of microorganisms.

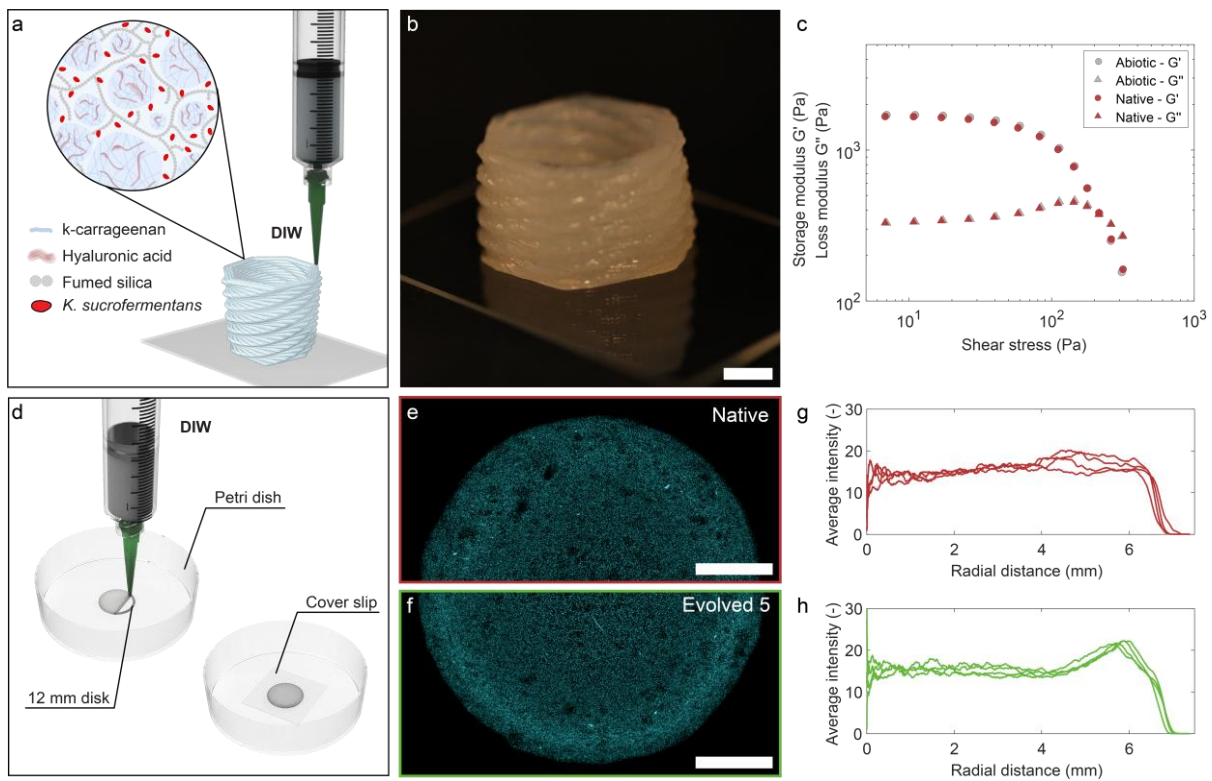

**Figure 4. 3D printed engineered living material. a.** Scheme of the Direct Ink Writing (DIW) process used to 3D print a complex-shaped object from a gel loaded with *K. sucrofermentans*. The gel contained 1.5 wt% κ-carrageenan, 1.5 wt% hyaluronic acid, 1.5 wt% fumed silica, growth media, and 0.36 M CFU/g bacteria. **b.** Image of the centimeter-scale engineered living object obtained by 3D printing the bacteria-laden ink. Scale bar: 1 cm. **c.** Rheological properties of the ink with no bacteria (grey) or with the native bacteria (red). The crossover between the storage (*G'*) and loss (*G'*) moduli occurs at a yield stress of 200 Pa for both inks. **d.** Schematics depicting the 3D-printed monolayer disks used to measure cellulose formation within the ink. A coverslip was utilized to limit oxygen access to the sides of the sample. **e-f.** Representative stitched confocal images of fluorescently labeled bacterial cellulose in the 3D-printed disks containing (e) the native or (f) the evolved (Ev5) strains after 1 day of incubation. Scale bars: 3 mm. **g-h.** Radial integration of the cellulose fluorescence over the whole disks containing the evolved and native strains ($n = 4$).

The biological machinery that controls the self-assembly of cellulose fibers in *K. sucrofermentans* can be affected by mutations in the bacterial genome or by metabolic changes arising from environmental stresses. To elucidate the origin of enhanced cellulose production in the evolved strains, we compared the genome of the evolved bacteria with that of the native strain. For this, a high-quality reference genome was obtained, using both shotgun and long-read methods to sequence the native strain. The assembled genome contained a single 2.95 Mbp genome containing 4 separate operons with cellulose synthase genes, and 4 plasmids (Figure 5a, Figure S12, Supporting Information), with a mean read depth of 112x (Table S1, Supporting Information). The genomes of the evolved and control strains were sequenced with the Illumina platform and aligned to the reference genome. The breseq algorithm [39] was used to predict mutations from annotated sequences.

Genomic analysis revealed no mutations in the control strain, whereas the 4 evolved strains identified as cellulose overproducers (Ev2- Ev5) shared a consistent and unique mutation: a 12-base pair deletion within the reading frame of the *clpA* gene (Figure 5a). Surprisingly, the deleted base pairs are not part of the *bcs* genes that code for the cellulose synthase complex sub-units (Figure 1a). Instead, these missing base pairs code for 4 adjacent amino acids from the N-domain of the ClpA protein (ClpA$_{84-87}$: QRVI, Figure 5a,b).

The ClpA protein is part of the ClpAPS protease complex, which is responsible for the hydrolysis of misfolded and degraded proteins inside cells. [40,41] The function of ClpA is to bind, unfold, and lead proteins to ClpP, where the hydrolysis process takes place. Under normal conditions, some of the proteins to be degraded are fixed through the specific interaction between the N domain of ClpA and the adaptor protein ClpS, as previously observed in *E. coli*. [42,43] The base pairs deleted from the genome of the evolved strain code for 4 amino acids of the ClpS-binding domain of ClpA. In particular, the Arginine residue ($R_{85}$) was shown to be highly conserved across species and involved in the interaction between ClpS and ClpA. [41]

We thus hypothesized that the in-frame 4-amino acid deletion in ClpA limited the ClpS-ClpA interaction without hindering the function of ClpA, thereby contributing to the increased cellulose production by the evolved bacteria (Figure 5b). To test this hypothesis, a *ΔclpS* knockout strain was generated from the native strain through homologous recombination using a pUC19 plasmid backbone. [44] Gel electrophoresis and DNA sequencing confirmed the successful deletion of the *clpS* gene in the knockout strain (Figure S13, Supporting Information). Experiments were performed to compare the *ΔclpS* knockout and native strains in terms of cellulose formation in bulk pellicles as well as bacterial growth in liquid culture medium.

The *ΔclpS* knockout strain exhibited enhanced cellulose production of 17% compared to the native strain (Figure 5c). The observed increase in cellulose production by the *ΔclpS* strain suggests a thus far unknown connection between the ClpS-ClpA interaction and cellulose regulation. ClpS has previously been shown to alter the degrading specificity of the ClpAP complex. [43] Our results suggest that such a shift in protein degradation specificity might favor cellulose production in *K. sucrofermentans*. This is in agreement with earlier research on *C. glutamicum*, which showed that a *clpS* deletion can increase the expression of a heterologous protein by 65%. [45] In addition to cellulose production, cell growth experiments revealed that the control, evolved, and *ΔclpS* strains had accelerated growth rates compared to the native strain (Figure 5d and Figure S14, Supporting Information).

The fact that the *ΔclpS* knockout strain did not reach the 37% increase in cellulose production observed in the evolved overproducers in this experiment indicates that a complete lack of ClpS-ClpA interaction might not be optimal, or that other factors than the ClpS-ClpA interaction might also contribute to the increased cellulose biosynthesis in the evolved bacteria. Since the control strain showed an accelerated growth rate (Figure 5d) and also produced 19% more cellulose than the native strain (Figure 3g), we expect the stresses imposed on the bacteria during the mutagenesis treatment steps to account for at least part of the non-genetic contributions to the enhanced cellulose production. We also noticed a 17% increase in cellulose production in sorted native cells compared to native unsorted cells (Figure S15a, Supporting Information), further indicating the role of epigenetic factors. Such an increase in cellulose production associated with sorting decreased to 3.6% when the sorting step was applied to the control strain (Figure S15b, Supporting Information). These findings highlight the combined importance of genetic diversification, environmental factors, and sorting steps in the directed evolution process to achieve the maximal performance of *K. sucrofermentans* in cellulose production.

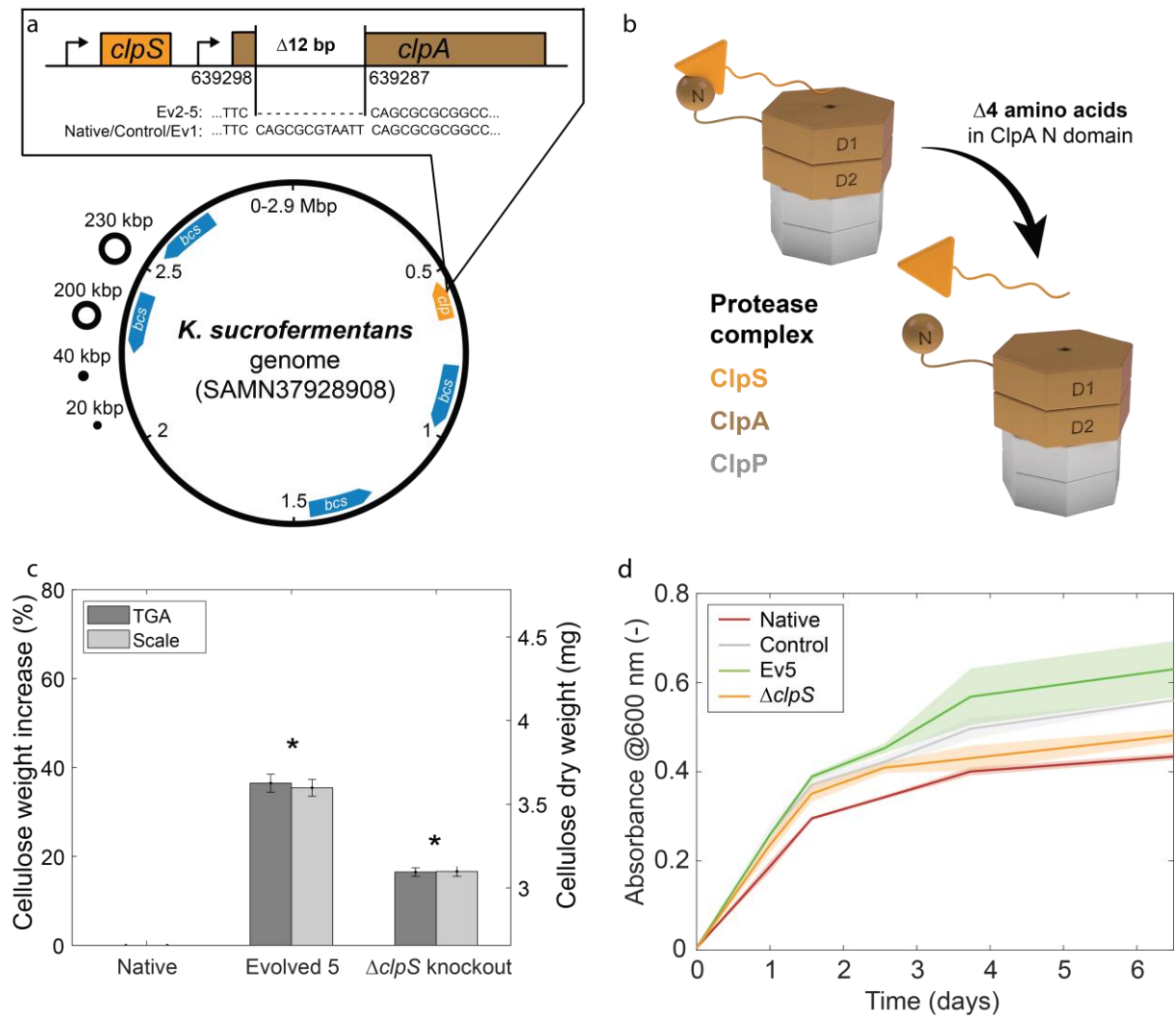

**Figure 5. Genomic analysis and mechanistic validation of gene-trait link. a.** Genome representation of the native strain (SAMN37928908, not to scale). Compared to the native bacteria, the evolved strains Ev2-5 display a 12-bp deletion in *clpA*, a gene coding for a protein involved in the ClpAPS protease complex. **b.** At the protein level, the genomic deletion translates into a 4-amino acid deletion in ClpA at the binding site for ClpS. Assuming that this ClpS-ClpA binding is impacted by the deletion, a *ΔclpS* knockout strain was engineered to validate this mechanism. The cartoon was adapted from Kim *et al.* [42] **c.** TGA and scale weight measurements of washed and air-dried bacterial cellulose pellicles (Ø = 3 cm, 5 mL growth media) produced by the native, evolved (Ev5), and *ΔclpS* knockout strains after 12 days of incubation from single colonies. Error bars correspond to the propagated error for the percentage data. Both the evolved and *ΔclpS* knockout strains showed cellulose production 37% and 17% higher compared to the native strain, respectively. Statistical analysis indicates a significantly different amount of cellulose produced by the different strains (*$p < 0.05$, $n = 5$). **d.** Bacterial growth curves of the different strains in the presence of 2 vol% cellulase in shaking conditions. Shaded areas around the curves represent the standard deviation ($n = 3$).

**Conclusions**

The directed evolution of the microorganism *K. sucrofermentans* in microfluidic droplets allowed for the discovery of a novel bacterial strain for the efficient production of cellulose in engineered living materials. A commonly used fluorescent dye was found to be an effective probe for the high-speed quantification of cellulose produced by single microorganisms inside the microfluidic droplets. Starting with a library of 40'000 randomly mutated variants, 4 microorganisms were identified as cellulose overproducers at the end of the directed evolution process. When cultivated for 12 days in culture medium, these selected bacteria were able to form pellicles with 37-70% more cellulose than the native strain. Using hydrogel inks as hosts for the selected bacteria, we show that the evolved microorganisms can be 3D printed into complex cellulose-based objects with structural features spanning multiple length scales. Genomic analysis revealed that the cellulose overproducers contained a 12-base pair deletion in the *clpA* gene, which encodes an important binding domain of the protease complex ClpAPS. The link between cellulose overproduction and the specificity of the ClpAPS complex was confirmed by the increased cellulose-forming ability of an engineered strain with dysfunctional protease units. By enabling the evolution of entire microorganisms towards specific functionalities, the high-throughput directed evolution platform shown in this work holds great potential as a discovery tool for novel microbial strains and genotype-phenotype links in engineered living materials. Moreover, the association between *clp* genes and cellulose overproduction opens new scientific questions and technological avenues for the genetic manipulation of cellulose-forming microorganisms. Beyond the realm of living materials, the directed evolution of whole microorganisms might also be an effective strategy to enhance the efficiency of existing and prospective biotechnological processes.

**Author Contributions**

A.S. conceptualized the research project with the support of J.L.. J.L., A.K., A.J., M.S., N.E., S.S., A.dM., and A.S. designed the experiments, and J.L., A.K., and A.J. conducted the experiments. J.L. performed mutagenesis, optimized the emulsifier, developed the fluorescent assay, characterized the droplets encapsulating bacteria, and analyzed the results of those steps. A.J. developed the sorter, conducted the screening and sorting experiments, and analyzed the corresponding results. J.L. performed all experiments quantifying cellulose production in bulk and in 3D-printed objects. A.K. performed the genome assembly and its analysis, and engineered the knockout strain. J.L. analyzed cell growth. J.L. and A.S. wrote the manuscript and prepared the figures. All authors discussed the results and contributed to the final version of the manuscript.

**Acknowledgments**

We thank ETH Zürich and the Swiss National Science Foundation (grant 200020_204614) for the financial support of this work. The research also benefitted from support from the Swiss National Science Foundation within the framework of the National Center of Competence in Research for Bio-Inspired Materials. We also thank our students Sami Moussalem and Aina Perappadan for their help with TGA measurements and analysis.

**Materials and Methods**

Cell culture for bacterial cellulose (BC) production

*Komagataeibacter sucrofermentans* JCM 9730 (ATCC 700178) is the cellulose-producing strain used in this study. Their growth media was composed of 25 g/L D-mannitol (Thermo Fisher Scientific), 5 g/L yeast extract (Sigma-Aldrich), 3 g/L peptone (Sigma-Aldrich), and optionally 15 g/L agar (Sigma-Aldrich) for solid medium. Frozen stocks of *K. sucrofermentans* (-80°C) were streaked on media plates to isolate single colonies, which were then inoculated in 5 mL growth media in 50 mL Falcon tubes (TPP). The lid was replaced with a foam plug to allow optimal oxygen availability. Cultures were incubated in static conditions at 28°C for 8-12 days to form bacterial cellulose (BC) pellicles at the air-media interface. For passage experiments (phenotype retention), a sterile inoculation loop was used to rub the BC pellicles and was streaked on solid medium, from which single colonies were picked for subsequent liquid cultures.

Cell concentration

All experiments started from frozen stocks of bacteria. For each stock, serial dilutions from $10^{-1}$ to $10^{-6}$ were prepared and plated on solid media in drops of 5 µL ($n = 6$). Colony-forming units (CFU) were counted at an appropriate dilution and averaged, and the stock concentration was back-calculated.

Absorbance measurements

Absorbance measurements were taken with the Varioscan LUX (Thermo Fisher Scientific) at 600 nm on 200 µL samples in 96-well plates (flat bottom, TPP) unless stated otherwise. Samples were always measured in triplicates, averaged, and blanks were subtracted. Blanks corresponded to media, with 2 vol% cellulase (Trichoderma reesei ATCC 26921, Sigma-Aldrich) and 35 µg/µL chloramphenicol (Sigma-Aldrich) in case it was also present in the samples.

UV-C mutagenesis

*K. sucrofermentans* from frozen stocks were inoculated in 150 mL of liquid media for 6 days at 28°C at 200 rpm. 2 vol% cellulase was added to the culture to digest any cellulose produced. After measuring the absorbance, cells were spun down at 3275 rcf for 10 min (Z306 Hermle) and resuspended in 0.9% NaCl (VWR) to reach a theoretical absorbance of 1. 10 mL of the cell suspension was added to each Petri dish (diameter Ø of 10 cm, TPP) to be exposed to different UV-C doses without the lid: 0, 0.5, 1, 2, 3, 5, 10, and 100 mJ/cm$^2$ (254 nm, UVP Crosslinker CL-3000, AnalytikJena). The plates were then left in the dark for 1 h, spun down, resuspended in enriched media (2X concentration) with 2 vol% cellulase, and incubated at 28°C and 200 rpm for either 1 h or 50 h. To store the cultures at -80°C for further analysis, aliquots with 20 vol% glycerol (Fisher BioReagents) were prepared. To quantify the cell survival, the 1 h cultures were serial diluted and plated on solid media (5 µL dots, $n = 6$). Colony-forming units (CFU) were counted at the $10^{-4}$ dilution and compared to the control sample, which was not exposed to UV-C (0 mJ/cm$^2$). The 50 h aliquots were used for the directed evolution process, assuming more recovery time was beneficial to increase our chances of finding a strain that grows sufficiently well in our culture media. The sample exposed to 0 mJ/cm$^2$ and recovered for 50 h was used as our control strain for the rest of the study to check the effect of the stresses the cells were exposed to (salt, dark, recovery) on cellulose production.

Fabrication of microfluidic devices

The photoresist SU-8 (3000 series, MicroChem) was patterned on silicon wafers using standard photolithography methods and used as masters for 1:10 PDMS (Sylgard$^{TM}$ 184, Dow Corning) soft lithography. Briefly, PDMS, after mixing, was poured on the masters, degassed, and cured at 70°C for 3-5 hours. After curing, devices were peeled off, and the inlet and outlet holes were punched. Both the emulsifier and sorter were bonded using an air-plasma treatment. Emulsifier devices were bonded to glass slides (Fisherbrand$^{TH}$ Superfrost$^{TM}$) using a 20 s air-plasma treatment at $4*10^{-1}$ mbar on medium level (Plasma Cleaner PDC-32G), and placed on a hot plate at 90°C for 1 h to promote further the bonding. Sorters were bonded to a PDMS-coated silicon wafer (spin-coated at 2400 rpm for 10 s and cured) through an air-plasma treatment for 1 min (Zepto, Diener), placed on a hot plate at 120°C for 2 h, and finally cut-off and bonded to a 50x24x0.17 mm glass coverslips (Glaswarenfabrik Karl Hecht), using the same air-plasma and heat treatment. To hydrophobize the channels of both devices, a 2 vol% solution of 1H,1H,2H,2H-perfluorooctyltrichlorosilane (Fluorochem) in HFE-7500 (3M) was flushed through the inlet and then air dried. For the sorting devices, electrodes were created by inserting a low melting point solder (51In/32.5Bi/16.5Sn; Indium Corporation) at one end of the electrode channel and a wire at the other end. The entire chip was subsequently placed on a 150°C hotplate for 3 minutes to melt and reflow the solder across the channel.

Single-cell encapsulation and incubation

*Cell loading:* Cell loading was determined based on the well-known Poisson distribution, according to which the probability ($p$) of a droplet to contain $k$ cells depends on the average number of cells per droplet volume λ as follows: $p(k,\lambda) = \frac{\lambda^k e^{-\lambda}}{k!}$. Knowing the droplet size and the frozen stock concentration, we aimed at a λ of 0.1 CFU/droplet, corresponding to a ~9.5% droplet occupancy. This low number was chosen to minimize co-encapsulation events (<0.5%). For each encapsulation experiment, cells were thawed from frozen stocks, and their concentration was adjusted to a λ value of 0.1 CFU/droplet (~1.6M CFU/mL). The occupancy was validated by calculating the percentage of droplets in which cellulose was produced through image analysis (see section Droplet image analysis).

*Cell encapsulation:* Cells were encapsulated in 49 µm droplets using a step emulsification PDMS microfluidic device. [46,47] The inner aqueous phase was composed of media, cells at low concentration (λ ~0.1 CFU/droplet), 218 µM Fluorescent Brightener 28 (FB, Sigma-Aldrich) to specifically stain the cellulose, and 26 µM fluorescein (Sigma-Aldrich) to generate a baseline signal in the droplets. The outer phase comprised 2 wt% 008-FluoroSurfactant in HFE-7500 (RAN Biotechnologies). Both phases were flown at 500 µL/h for 10 minutes, which enabled the formation of more than a million droplets.

*Droplet incubation:* Droplets were incubated in hydrophobic glass vials at 28°C on top of 200 µL HFE-7500 (3M) in static conditions for 24 h before the sorting process. Vials were horizontally placed to form a monolayer of droplets and thereby ensure that bacteria had equal access to oxygen during incubation.

Droplet image analysis

Prior to screening in microfluidic devices, droplet sizes and cellulose content were analyzed using z-stacks of confocal images (TCS SP8, Leica, 30 slices, 3.58 µm each). A 405 nm laser was used to excite the cellulose-specific dye (Fluorescent Brightener 28 (FB), Sigma-Aldrich), which was detected between 432 nm and 460 nm (HyD detector). An additional PMT transmission detector was used to image the droplets. Using ImageJ,[48] FB images were summed, and one of the droplet images was chosen for edge detection. Droplet diameters were estimated using the Hough Transform plugin (UCB Vision Sciences) after image thresholding. For fluorescence quantification in the droplets, the FB sum stack and chosen droplet images were imported to Cell Profiler 4.1.3,[49] a user-friendly software previously reported for droplet image analysis.[50] Briefly, droplets were detected as objects in a desirable size range and filtered out if on the edge of the image before measuring the intensity of each object. Results were exported as an Excel file, and all were analyzed and plotted with MATLAB (R2023a, Math Works). Droplets were considered occupied when their measured FB fluorescence exceeded the maximum fluorescence detected on Day 0.

Droplet sorting and screening

*Reinjection:* The incubated droplets were reinjected into the droplet sorting device by pressurizing the incubation glass vial at 100-150 mbar (LineUP Flow EZ, Fluigent), transferring the droplets from the vial to the inlet of the sorting device via ID/OD 0.86/1.32 mm polyethylene tubing (Scientific Commodities). Oil (HFE-7500, 3M) was delivered at a flow rate of 6 -8 µL/min on both sides of the microfluidic channel to increase the spacing between the flowing droplets.

*Droplet sorting setup:* The incoming droplets were excited using a 405 nm solid-state laser (Omicron Laserage Laserproduckte) beam (expanded, shaped into a line, and focussed on the chip using a 20x/0.45 NA Nikon S Plan Fluor ELWD objective). The fluorescence emission of these droplets was first filtered through a 30 µm pinhole and a 480/30 optical filter (Chroma Technology) and was finally captured using a PMT (H10722-20; Hamamatsu Photonics). The voltage signal from the PMT was sampled at a rate of 100,000 samples per second by an FPGA (NI PXI-7842R; National Instruments) running a custom LabVIEW code. The code registered the signal from incoming droplets and, based on a user-defined threshold, issued a series of tunable pulses (25-35 pulses at a pulse frequency of 40 kHz, corresponding to a train length of 0.6-0.8 ms) that were fed to the chip via high-voltage amplifier (Trek 632B, Advanced Energy) for droplet sorting (900 V final pulse voltage).

*Mutant library sorting:* The added fluorescein in the droplets provided a fluorescence baseline to detect all droplets and was adjusted to emit a signal corresponding to 0.05 V. A total of 430'000 droplets were screened in less than 10 min, of which 500 were sorted using a threshold of 2.75 V. Those droplets were collected in a sterile Eppendorf tube. After adding media and demulsifying, the droplets were plated onto solid media. Five evolved single colonies were then arbitrarily picked and grown in liquid media with 2 vol% cellulase, and aliquots with 20 vol% glycerol (Fisher BioReagents) were frozen for further analysis (Ev1-Ev5).

*Droplet screening:* To compare the native, control, and evolved variants, the fluorescein baseline was adjusted to 0.1 V across all the measurements. 15'000 droplets of each variant were screened. The screening data was fitted to a Gaussian mixture distribution model with 3 components and baseline-adjusted using the lowest peak's mean (corresponding to the fluorescein signal) using the following formula: $\frac{data*0.1}{fluorescein\ mean\ signal}$. The middle mean value (fluorescence signal from the Fluorescent Brightener 28) reporting the cellulose production in the majority population was used for comparison among the native, control and evolved strains.

Bacterial cellulose (BC) dry weight

To quantify the amount of BC produced by the evolved strains compared to the native and control (0 mJ/cm$^2$) strains, triplicates of BC pellicles were grown, as previously explained. After 12 days, the pellicles were washed 3 times with 0.1 M NaOH (Fisher Scientific) at 60°C in a water bath over a period of 24 h and then brought back to neutral pH washing 3 times with MilliQ water (NANOpure Diamond, Barnstead). They were then dried in a 60°C oven for 48h on Teflon films (McMater-Carr) to avoid sticking and stored under vacuum until analyzed. Dried bacterial cellulose pellicles were weighed with a precision balance (UMT2 Microbalance, Mettler Toledo). Thermogravimetric analysis (Discovery TGA 5500, TA Instruments) was also performed on the dried cellulose pellicles by increasing the temperature to 650°C at a rate of 10°C/min. During TGA, an isotherm of 15 min at 120°C was applied to remove all the humidity contained in the samples. Weight loss was then calculated between the end of the 120°C isotherm and the end of the program at 650°C.

3D printing

*Ink preparation*: The ink was composed of 1.5 wt% sodium hyaluronate (BulkSupplements), 1.5 wt% κ-carrageenan (Acros Organics), and 1.5 wt% fumed silica (WDK V15, Wacker Chemie) in the standard media used for *K. sucrofermentans* described above. The ink was prepared following a previously published protocol.[3] UV-C was used to sterilize all powders before mixing. 218 μM of Fluorescent Brightener (Sigma-Aldrich) was also added to stain the cellulose produced in the ink. For each experiment, a new batch of ink was prepared and separated in three equal amounts, in which 0.36M CFU/g of bacteria were added from frozen stocks. For controls, the same volume of sterile media was added to the ink (900 µL). The final inks were loaded into 10 mL syringes and kept at 4°C until 3D printing on the same day.

*Ink rheology*: Rheological measurements of the inks were performed under oscillatory and steady-shear conditions using a sandblasted parallel plate geometry (PP25-S, Anton Paar) at 25°C (MCR 302 compact rheometer, Anton Paar). The storage and loss moduli (*G'* and *G''*) were measured via oscillatory rheology by applying a shear strain amplitude increasing logarithmically from 0.01 to 100% at a constant frequency of 1 s$^{-1}$. Steady-shear flow curves were then recorded under strain-controlled conditions by ramping the shear rate logarithmically from 0.01 to 100 s$^{-1}$ and then back from 100 to 0.01 s$^{-1}$.

*Direct ink writing (DIW)*: All structures were 3D printed via the DIW technique using a 10 mL syringe, a 0.84 mm needle, and a layer height of 0.7 mm. For the quantification of the cellulose content grown within the ink, 1-layer disks of 12 mm diameter were printed directly on Petri dishes. Immediately after completion of the print, the disks were covered with coverslips to ensure that oxygen was only available from the sides. Each Petri dish was then sealed with parafilm and imaged as is with a confocal microscope (TCS SP8, Leica) after a day of incubation at 28°C and 85% relative humidity in static conditions. For each full disk, a 15-slice stack of 750 µm total was imaged (excitation: 405 nm; emission: 432-460 nm) and stitched together using the Leica software.

*Image analysis*: All the image analysis was performed using ImageJ. [48] Slices of the stacks were averaged, and the radial profile of each disk was plotted. Data was then smoothed and plotted with MATLAB (R2023a, Math Works).

Genome sequencing, assembly, and analysis

Genome sequencing was performed by MicrobesNG, United Kingdom. The *K. sucrofermentans* JCM 9730 (ATCC 700178) reference genome was sequenced with an Illumina NovaSeq 6000 (Illumina, San Diego, USA) using a 250 bp paired-end protocol, as well as with GridION (Oxford Nanopore Technologies, UK) to obtain long reads. Strains selected from the directed evolution process were sequenced with lllumina NovaSeq 6000 (Illumina, San Diego, USA) only.

Reference genome assembly combined both long and short reads using Unicycler version 0.4.0 [51], and annotation was performed with Prokka 1.13. [52] To quantify the quality of the assembly, coverage statistics were calculated using the short read data using samtools 1.3.1. [53] The obtained statistics are shown in Table S1 (Supporting Information). Mutations were detected using breseq 0.35.1, [39] aligning the Illumina paired-end reads to the annotated reference genome. Bioinformatic analysis of specific genes and operons was performed with BioPython. [54]

Knockout generation

*K. sucrofermentans* knockout was produced by transforming cells with a pUC19 plasmid, which does not replicate inside this bacterium. Further, the pUC19 plasmid was engineered to contain 500 bp regions of homology to the genome around the *clpS* sequence, designed to remove the start codon of *clpS* and disrupt expression (Figure S13b, Supporting Information). The homology arms were placed on each side of a *cat* chloramphenicol resistance cassette to create plasmid pUC19-clpS-KO (Figure S13a, Supporting Information) through Gibson assembly [55] and cloning in *E. coli* DH5a cells. DNA oligos were obtained from Integrated DNA Technologies (IDT, Belgium).

*K. sucrofermentans* cells were made electrocompetent by growing a culture to an absorbance of 0.8 at 600 nm, before centrifuging at 3000 g for 10 min and washing the pellets three times with 10% glycerol (Fisher BioReagents). Cells were then concentrated and electroporated with pUC19-clpS-KO. Following overnight recovery in media supplemented with 2 vol% cellulase at 28°C in shaking conditions (200 rpm), cells were plated onto 25 µg/ml chloramphenicol to select for colonies containing the *cat* cassette. Colonies were screened by colony PCR with GoTaq polymerase (M7423, Promega), using verification primers designed around the insertion site in *clpS* on the genome, *clpS_checkF*: CGAGCACCGCCTGCTCCACCG and *clpS_checkR*: TCGCGACGGGGGCGTTAAGATG. PCR products were run in 1.5% agarose (Ultrapure Agarose, Thermo Fisher Scientific) with gel electrophoresis in Tris-Acetate-EDTA (TAE) buffer (Fisher Bioreagents). Sybr Safe (Thermo Fisher Scientific) DNA stain was added initially to the gel, and after electrophoresis alongside a 1kb Plus DNA Ladder (Thermo Fisher Scientific), imaging was performed using a ChemiDoc™ MP imaging system (BioRad, Figure S13c, Supporting Information). Further verification was performed by Sanger Sequencing the PCR products with the verification primers (Microsynth, Switzerland), which in each case found the expected sequence.

Growth curve

Each tested strain was thawed, spun down at 3000 rcf for 10 min (5417R, Eppendorf), resuspended in 1 mL of fresh media, and their absorbance at 600 nm adjusted to 0.005 by dilution. 2 vol% of cellulase was added to each culture to prevent the formation of a cellulose pellicle. 1 mL of each culture was then transferred to 24-well plates (flat bottom, TPP) in triplicates, and the well plates were covered with a breathable film (BREATHseal™, Greiner Bio-One) to avoid contamination between wells. The well plates were incubated at 28°C and 85% relative humidity in shaking conditions (200 rpm). Every day, the breathable film was removed for the measurement, the absorbance at 600 nm of each well was measured, triplicates were averaged, and blanks composed of media and 2 vol% cellulase were subtracted from the measurement. The breathable film was changed, and the plates were incubated until the next measurement.

Statistical analysis

The significance of the increased cellulose production was evaluated with a one-way ANOVA (MATLAB R2023a, Math Works), followed by a pairwise comparison if the results showed a statistically significant difference between the groups ($p < 0.05$). A Bonferroni correction was applied to compensate for the effects of multiple comparisons.

Data availability

Genome sequences of the *K. sucrofermentans* native JCM 9730 (SAMN37928908), evolved JML 2321 (SAMN37928909), and *ΔclpS* knockout JML KO 23 (SAMN37928910) strains were uploaded to NCBI and will be available from [date of publication].

**Supporting Information**

**Table**

**Table S1.** Mean Illumina sequencing depths. To verify the quality of the sequencing, the mean sequencing depth was determined for each piece of DNA in the long-read reference genome and in each Illumina sequencing run, quantifying the mean number of reads for each nucleotide position.

|  | Reference Genome (Native) | Native | Control | Ev1 | Ev2 | Ev3 | Ev4 | Ev5 | Sorted Native | Sorted Control |
|---|---|---|---|---|---|---|---|---|---|---|
| **Genome** | 110.1 | 105.2 | 68.9 | 58.1 | 60.7 | 83.8 | 56.9 | 48.9 | 77.2 | 55.9 |
| **Plasmid p1** | 102.5 | 103.2 | 58.1 | 38.8 | 40.9 | 54.1 | 68.8 | 60.1 | 74.4 | 44.2 |
| **Plasmid p2** | 171.5 | 101.0 | 56.7 | 46.3 | 45.6 | 58.1 | 73.1 | 60.3 | 82.3 | 47.1 |
| **Plasmid p3** | 89.70 | 94.40 | 17.9 | 12.7 | 16.4 | 20.4 | 19.1 | 13.8 | 22.9 | 19.0 |
| **Plasmid p4** | 91.60 | 91.80 | 23.2 | 13.2 | 27.0 | 25.7 | 23.6 | 17.6 | 31.1 | 21.9 |

**Figures**

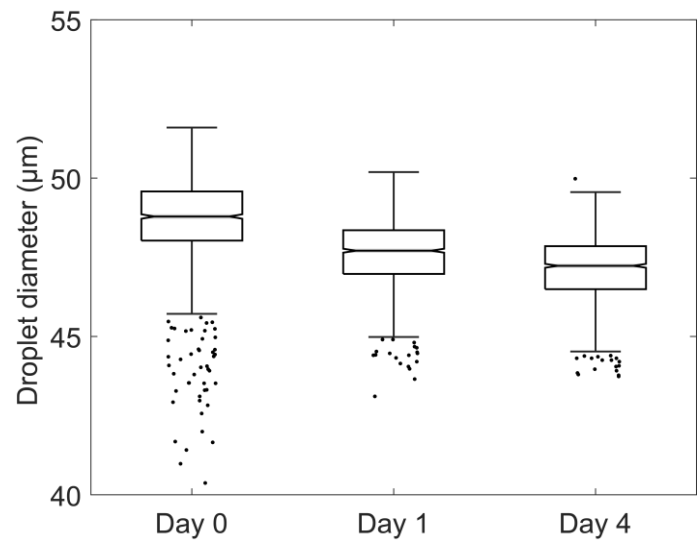

**Figure S1.** Diameters of cell-laden droplets (λ ~0.1 CFU/droplet) produced with the parallelized step emulsification microfluidic device. Statistical analysis led to the following mean droplet diameters (Ø) and polydispersity indices (PDI): Day 0: Ø = 48.8 μm, PDI = $1.1*10^{-3}$; Day 1: Ø = 47.7 μm, PDI = $4.9*10^{-4}$; Day 4: Ø = 47.2 μm, PDI = $4.8*10^{-4}$ ($n > 1000$).

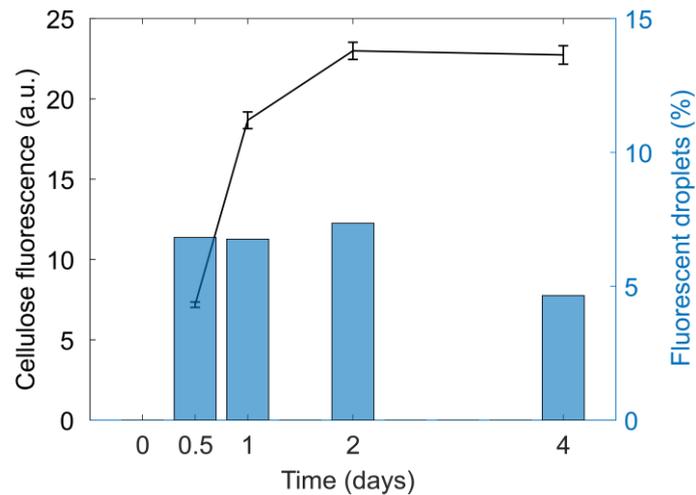

**Figure S2.** Evolution of the quantity of fluorescently labeled cellulose in droplets over time. Confocal microscopy images were analyzed, and the cellulose-correlated fluorescence in individual droplets was measured ($n$ ~1200 droplets). The majority of the cellulose was produced within the first day of droplet incubation, with a maximum fluorescence reached after two days of incubation (left axis). Error bars correspond to the standard error of the mean. About 7% of droplets contained cellulose-producing bacteria, which is close to the 9.5% value expected from Poisson statistics for a cell concentration λ of 0.1 CFU/droplet (right axis).

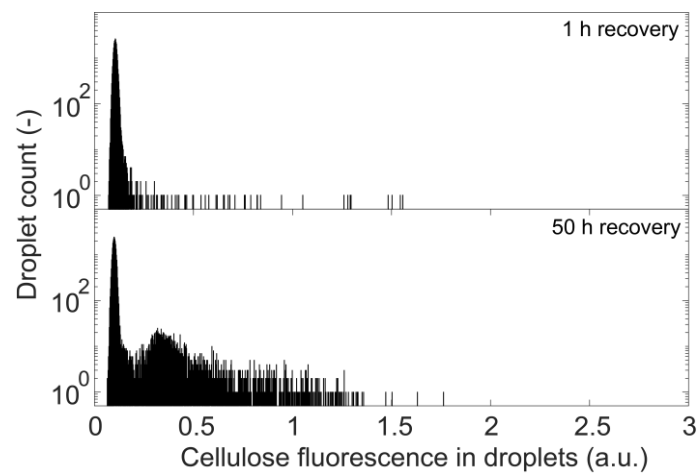

**Figure S3.** Histogram of cellulose fluorescence signals in droplets containing single cells after mutagenesis (10 mJ/cm$^2$) with 1 h or 50 h of recovery (λ ~0.1 CFU/droplet, 28'000 droplet events). Screening was performed after incubation of the droplets at 28°C for 24 h. Mutants recovered after 50 h show more cellulose than in the 1 h-recovered strain, as evidenced by the highly fluorescent droplets detected for the former case. Using the library of 50 h-recovered mutants is expected to increase the chances of finding a suitable candidate for growth under laboratory conditions.

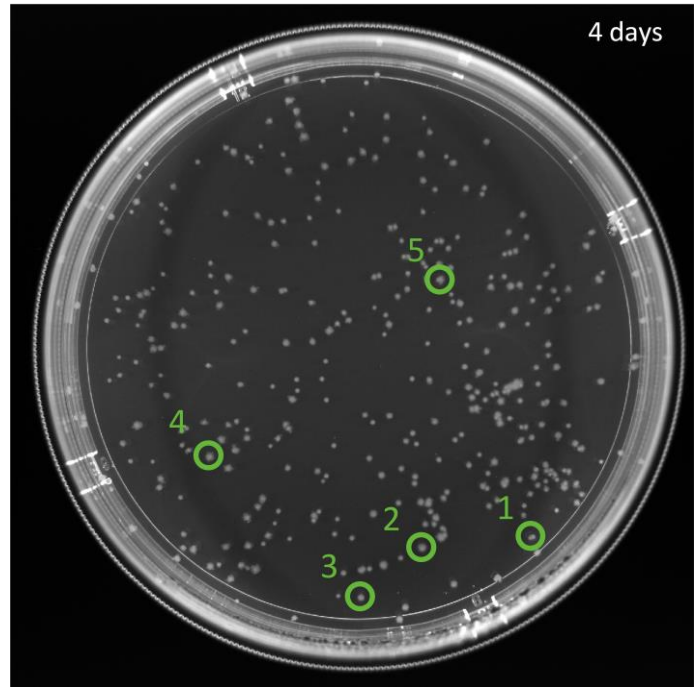

**Figure S4.** Grown colonies of selected mutants after 4 days. Sorted droplets containing evolved bacteria were collected and spread onto solid media to enable the growth of individual mutants into colonies. Five of them were chosen for further analysis (Ev1-Ev5, green circles).

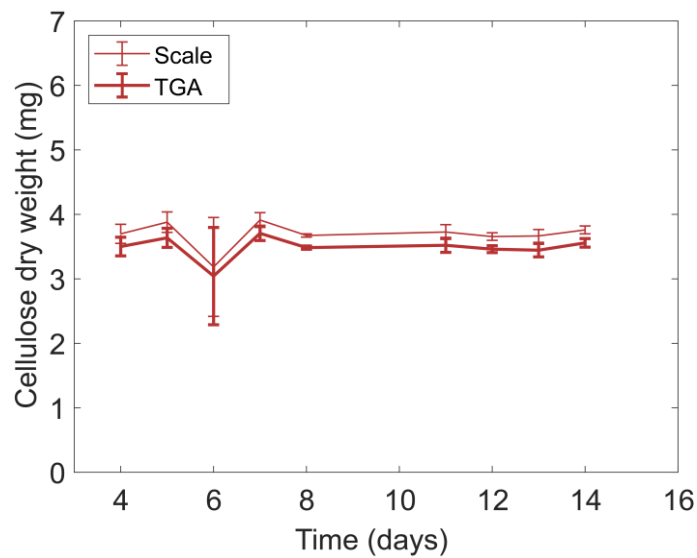

**Figure S5.** Weight of bacterial cellulose (BC) pellicles formed by the native strain over time. BC pellicles with a diameter of 3 cm were grown from single colonies under static conditions in 5 mL growth medium at 28°C. After growing for different amounts of days, the pellicles formed at the air-water interface were washed and air-dried to determine the variability in cellulose dry weight resulting from slightly different inoculation quantities. Weights were measured with both a laboratory scale and extracted from TGA for each washed and air-dried sample ($n$ = 3). After 8 days, the weight of the cellulose stabilized. Error bars represent standard deviations.

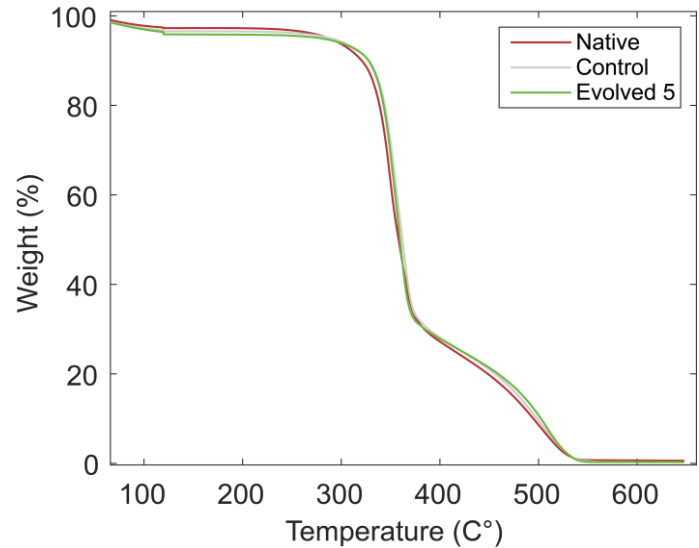

**Figure S6.** Representative thermogravimetric analysis (TGA) curves of bacterial cellulose produced by the native (red), control (grey), and evolved (green) strains after 12 days of incubation of single colonies at 28°C in static conditions. The washed and air-dried samples were heated up to 120°C (10°C/min), then an isothermal step was set for 15 min to ensure all the water evaporated, and finally, samples were thermally degraded to 650°C (at 10°C/min). Weight loss was then calculated between the end of the isotherm and the end of the program.

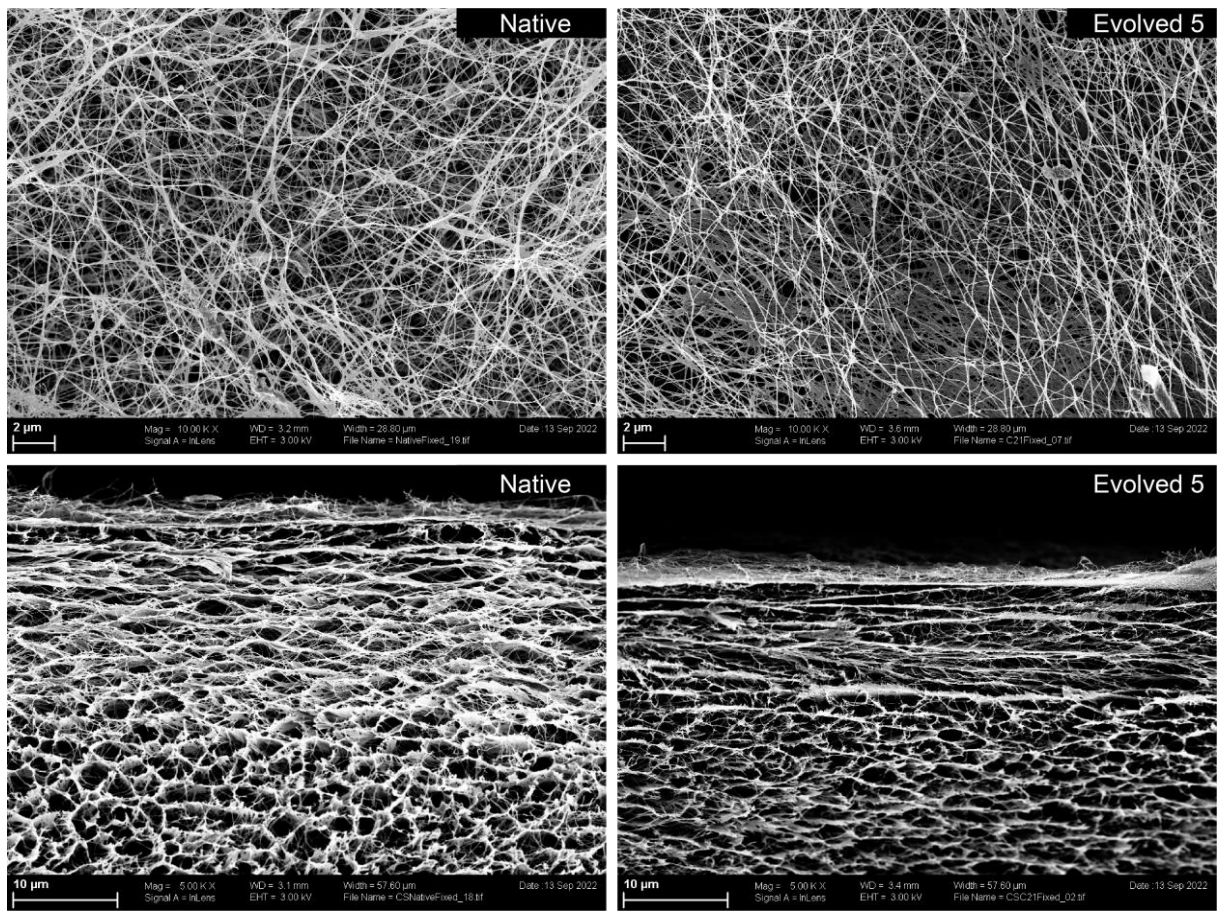

**Figure S7**. Scanning electron microscopy of freeze-dried bacterial cellulose pellicles produced by the native (**a, c**) and evolved 5 (**b, d**) strains after 12 days of incubation at 28°C in static conditions. **a** and **b** originate from the bottom fibers of the pellicle. **c** and **d** show the cross-section of the pellicle. The long fibers produced by both strains appear similar.

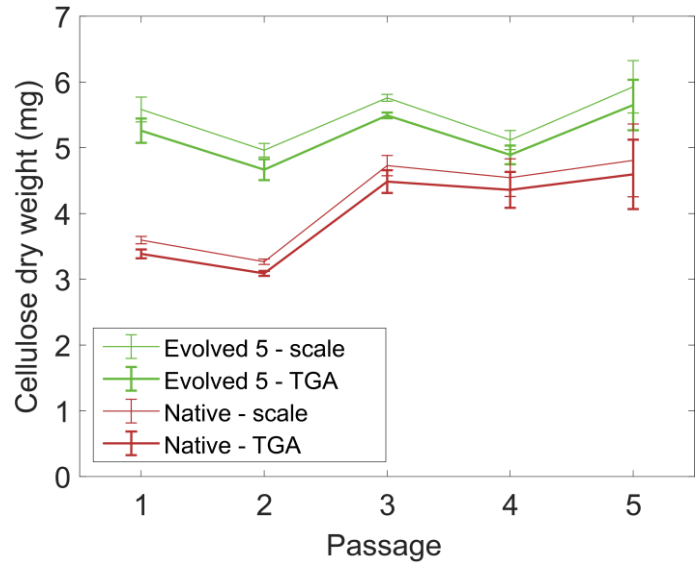

**Figure S8.** Weight of bacterial cellulose (BC) pellicles grown after increasing number of passages. BC pellicles with a diameter of 3 cm were grown under static conditions in 5 mL growth medium at 28°C from single colonies for 8-12 days. The single colonies were obtained from previous pellicles over 5 generations. The weight was measured both with a laboratory scale and extracted from TGA for each washed and air-dried sample ($n$ = 3). The evolved strain (green) shows a stable phenotype of increased cellulose production. The native strain (red) shows a spontaneous increase in cellulose production after the third generation, but always less than the evolved strain. Error bars represent the standard deviation.

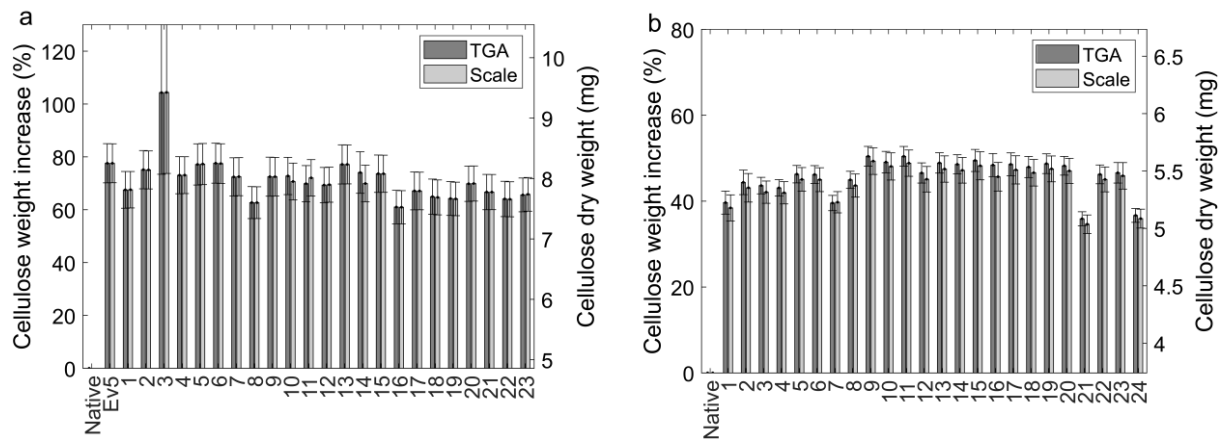

**Figure S9.** Second round of directed evolution of *K. sucrofermentans*. The evolved strain from the first round of directed evolution (Ev5) was mutated (10 mJ/cm$^2$), encapsulated (λ ~0.1 CFU/droplet), and sorted with two different voltage thresholds: the same as the first round 2.75V (**a**) and a higher threshold of 3.60V (**b**). Bacterial cellulose pellicles with a diameter of 3 cm were grown under static conditions for 12 days from single colonies of all the selected strains of this second round (5 mL growth media, 28°C, static). The weight was measured both with a laboratory scale and extracted from TGA for each washed and air-dried sample ($n$ = 3). All selected strains show a phenotype comparable to Ev5 but no further increase in cellulose production. Error bars represent the propagated errors for the percentage data.

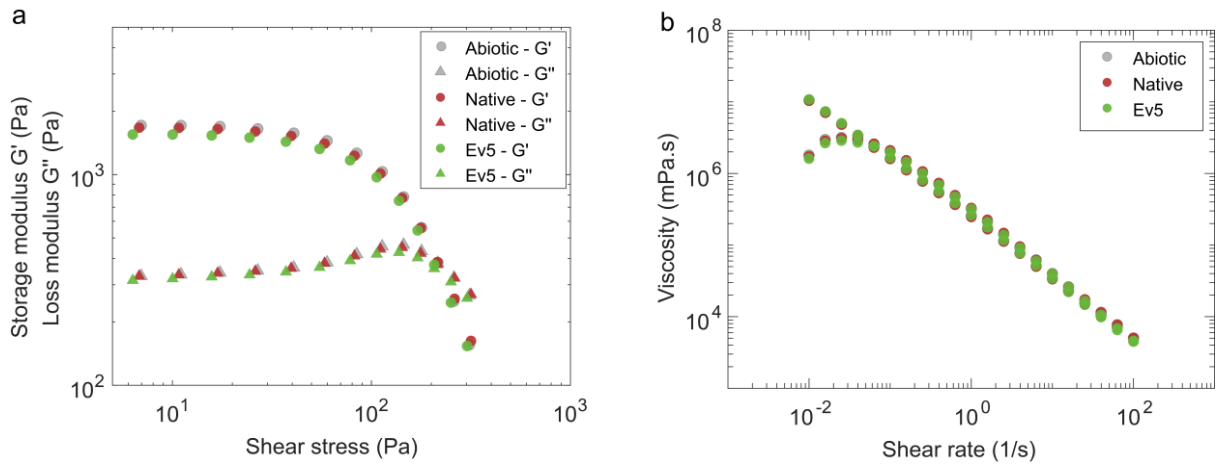

**Figure S10.** Rheological properties of the cell-laden gel used for 3D printing. **a.** Oscillatory rheology of gels containing native or evolved (Ev5) strains compared to the cell-free (abiotic) ink. The storage (*G'*) and loss (*G''*) moduli of the gels were measured for applied shear strains that varied from 0.01 to 100%. The gel exhibits an elastic response at low shear stresses (*G'* > *G''*) and becomes predominantly fluid (*G''* > *G'*) above 200 Pa. **b.** Steady-shear flow curves were obtained by applying a shear rate that increased from 0.01 to 100 s$^{-1}$. The gels show a shear-thinning behavior that is suitable for 3D printing via Direct Ink Writing. Grey: abiotic gel; red: gel with native bacteria (0.36 M CFU/g bacteria); green: gel with evolved bacteria 5 (0.36 M CFU/g bacteria).

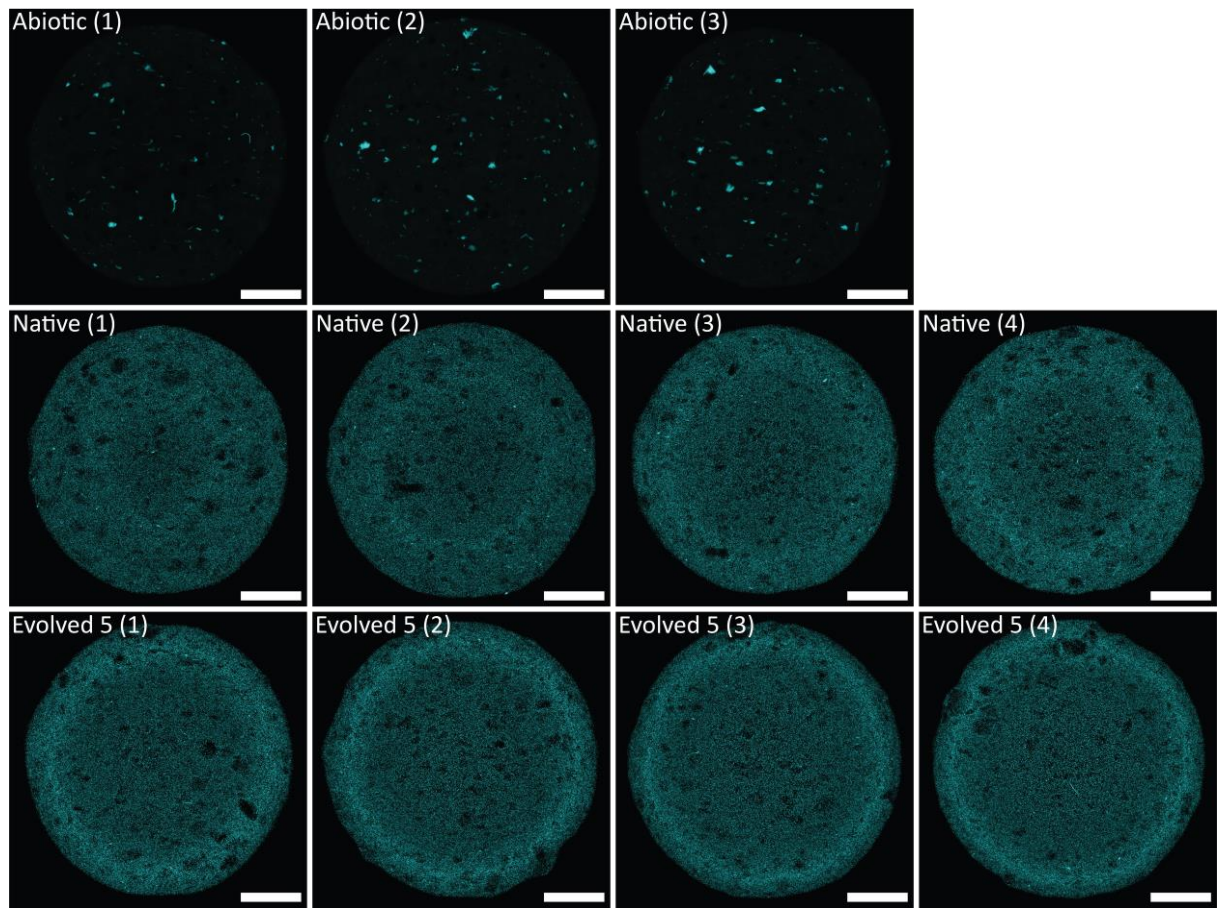

**Figure S11.** Confocal images of 3D-printed bacteria-laden disks (0.36 M CFU/g bacteria) after 1 day of incubation at 28°C (*n* = 4). The 12-mm disks are covered with a cover slip to ensure oxygen is only available through the sides. Fluorescently labeled cellulose is imaged using excitation and emission wavelengths of 405 nm and 432-460 nm, respectively. Images show that both strains produce more cellulose at the edge of the disk, at the air-water interface. The evolved strain 5, however, produces

more cellulose in a narrower ring, whereas the ring where the native strain produces more cellulose is broader and less intense. Scale bars: 3 mm.

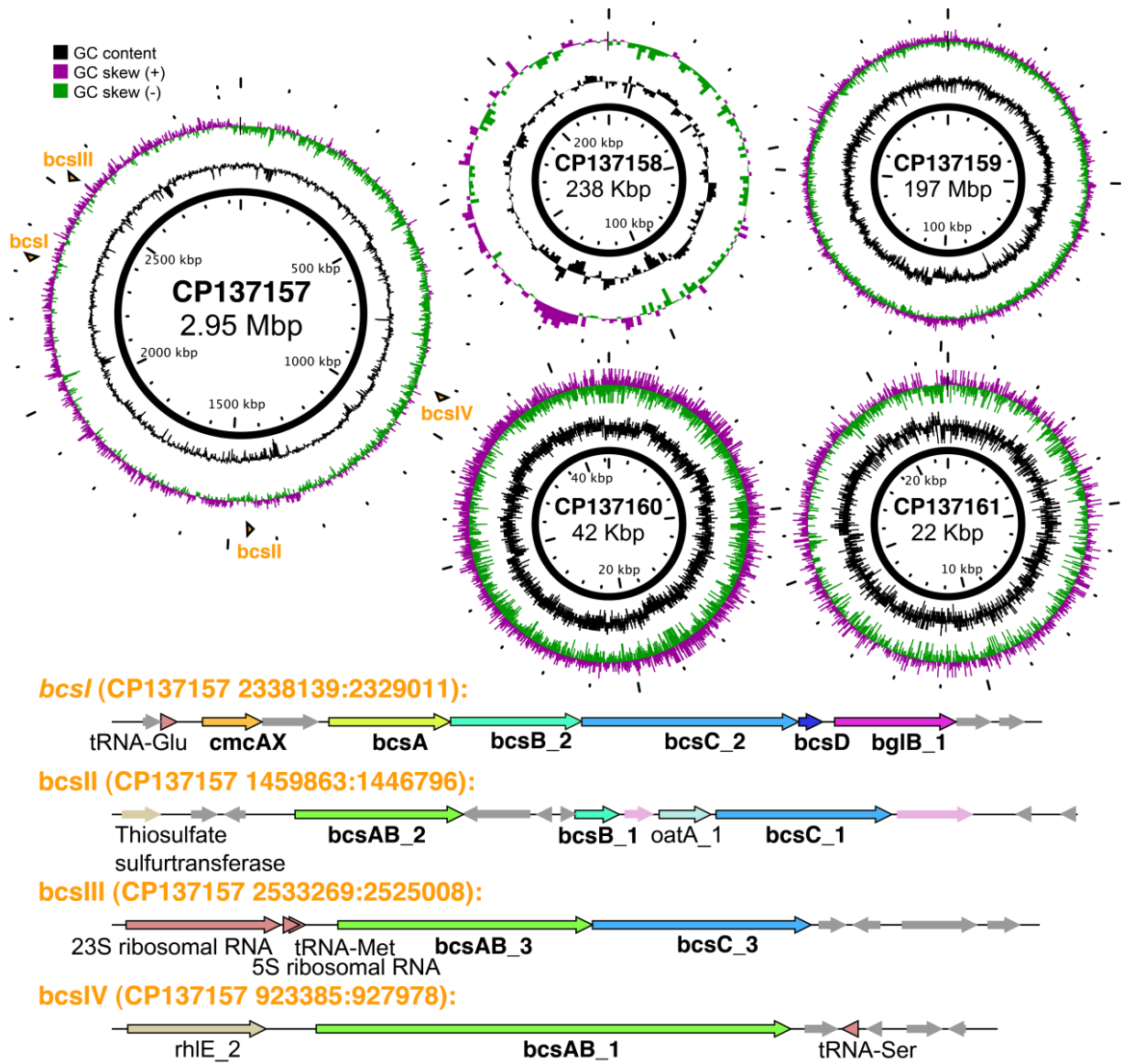

**Figure S12.** *Komagataeibacter sucrofermentans* assembled genome (NCBI accession SAMN37928908). The genomic DNA (CP137157) is 2.95 Mbp and contains 4 operons containing genes coding for bacterial cellulose synthase subunits. *bcsI* (2338139:2329011) contains *bcsA*, *bcsB_2*, *bcsC_2* (also annotated as *acsC_2*), and *bcsD* (*acsD*) genes. *bcsII* (1459863:1446796) contains *bcsAB_2* (*acsAB_2*), *bcsB_1*, and *bcsC_1* (*acsC_1*) genes. *bcs III* (2533269:2525008) contains *bcsAB_3 (acsAB_3)*, and *bcsC_3 (acsC_3)* genes. *bcsIV* (923385:927978) contains only the *bcsAB_1* (*acsAB_1*) gene. Four additional plasmids are present in this strain: p1 CP137158 (238 Kbp), p2 CP137159 (197 Kbp), p3 CP137160 (42 Kbp), and p4 CP137161 (22 Kbp).

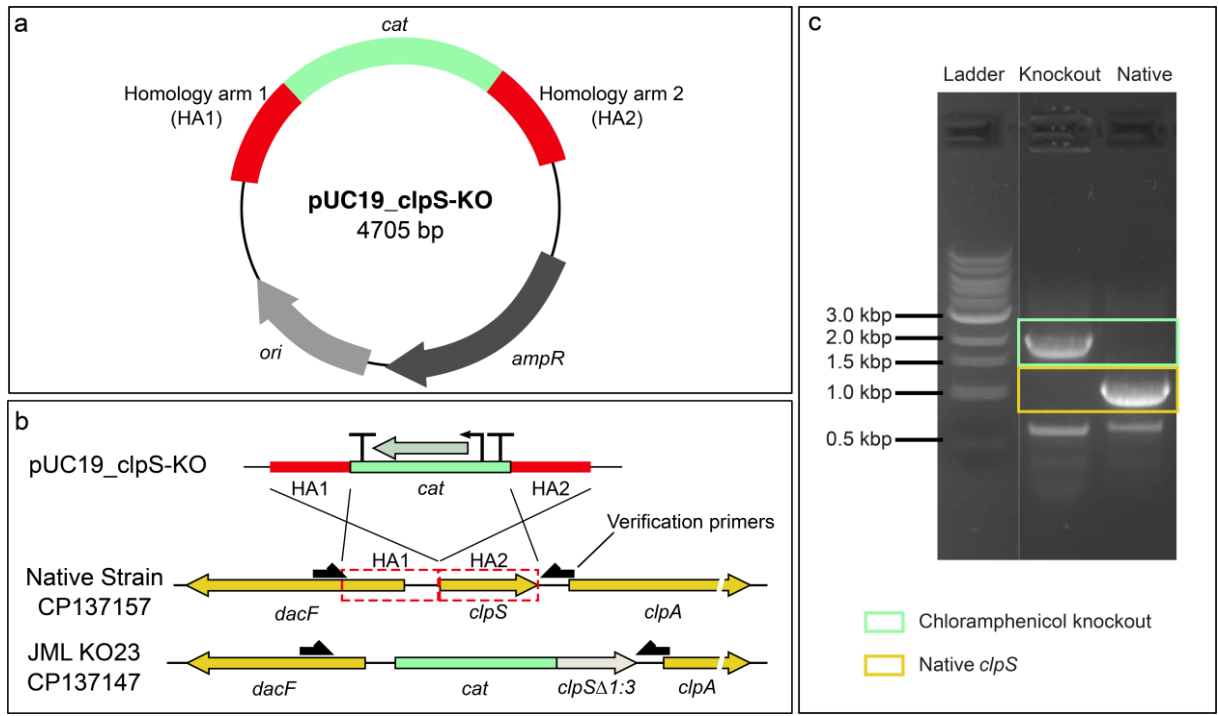

**Figure S13.** *ΔclpS* knockout generation. **a.** pUC19-clpS-KO plasmid engineered with a pUC19 ampicillin-resistant backbone. 500 bp regions of homology to the *K. sucrofermentans* native genome were designed around the *clpS* sequence (HA1 and HA2). Between the homology arms, a *cat* chloramphenicol resistance cassette was inserted. **b.** pUC19-clpS-KO is designed to remove the start codon of *clpS* and disrupt expression. Verification primers *clpS_checkF* and *clpS_checkR* were designed to confirm the knockout with PCR and sequencing (black arrows). **c.** Gel electrophoresis of PCR amplicons around *clpS* in *ΔclpS* knockout and native strains. The 2 Kpb band in the *ΔclpS* knockout strain confirms the insertion of a chloramphenicol resistance gene into the *clpS* sequence. The 1 Kbp band in the native strain shows the native *clpS* amplicon.

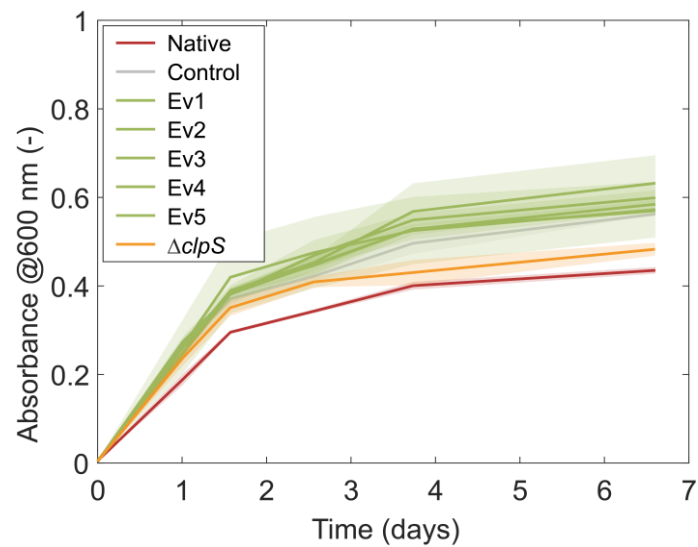

**Figure S14.** Bacterial growth curves of the different strains in the presence of 2 vol% cellulase in shaking conditions (200 rpm, 28°C, 85% relative humidity). To innoculate similar amounts of bacteria, the absorbance of the different samples was adjusted to 0.005 at 600 nm. Shaded areas around the curves represent the standard deviation ($n = 3$).

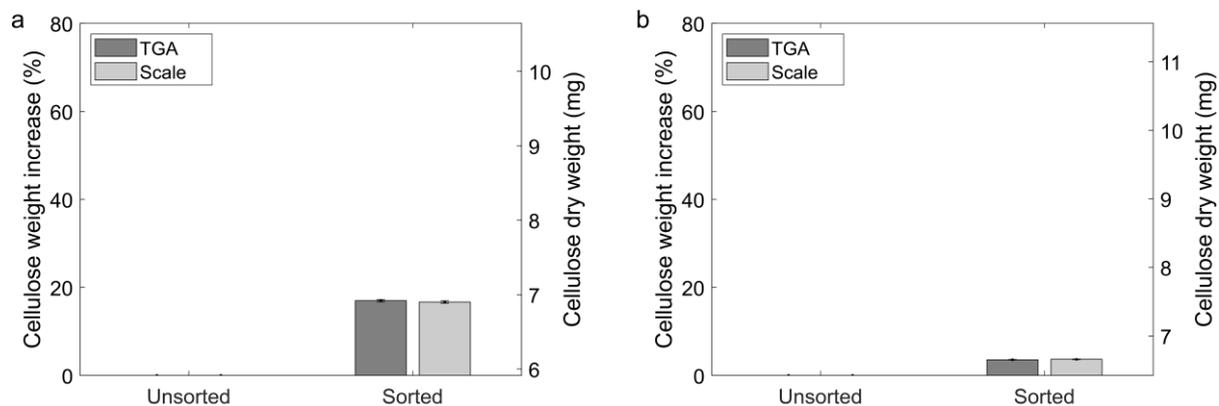

**Figure S15.** Weight of bacterial cellulose (BC) pellicles obtained from sorted and unsorted strains. The native (**a**) and control (**b**) bacteria were encapsulated using a cell suspension with concentration λ ~0.1 CFU/droplet and afterward sorted in the microfluidic device utilizing a PMT voltage of 2.75 V. Pellicles with a diameter of 3 cm were grown from single colonies of native unsorted and sorted samples (**a**) or control unsorted and sorted samples (**b**) under static conditions for 12 days in 5 mL growth media at 28°C. Weights were measured both with a laboratory scale and extracted from TGA for each washed and air-dried sample. Increases in cellulose production of 17% and 3.6% were observed when the native and control bacteria were sorted, respectively. Error bars represent the propagated errors for the percentage data ($n$ = 3).